\documentclass [12pt,a4paper] {article}

\usepackage {graphicx}
\usepackage {color}
\usepackage {amssymb}
\usepackage[affil-it]{authblk}	
\usepackage {slashbox}		

\begin {document}

\title {The emission of energetic electrons from the complex
streamer corona adjacent to leader stepping}

\author {Christoph K\"ohn$^{1}$, Olivier Chanrion$^{1}$, Kenichi Nishikawa$^{2}$, Leonid Babich$^{3}$, Torsten Neubert$^{1}$}

\affil{$^{1}$ Technical University of Denmark, National Space Institute (DTU
Space), Elektrovej 328, 2800 Kgs Lyngby, Denmark}
\affil{$^{2}$ Department of Physics, Alabama A\&M University, Normal,
AL35762, USA}
\affil{$^{3}$ Russian Federal Nuclear Center-VNIIEF, Sarov, Russia}

\date{}

\maketitle 

\begin {abstract}
We here propose a model to capture the complexity of the streamer corona
adjacent to leader stepping and relate it to the production of
energetic electrons serving as a source of X-rays and $\gamma$-rays,
manifesting in terrestrial gamma-ray flashes (TGFs). During its stepping, the leader tip is accompanied by a
corona consisting of multitudinous streamers perturbing the air in its vicinity and
leaving residual charge behind. We explore the relative importance of air perturbations and preionization on the production of
energetic run-away electrons by 2.5D cylindrical Monte Carlo particle simulations
of streamers in ambient fields of 16 kV cm$^{-1}$ and 50 kV cm$^{-1}$ at ground
pressure. We explore preionization levels between $10^{10}$ m$^{-3}$ and $10^{13}$ m$^{-3}$,
channel widths between 0.5 and 1.5 times the original streamer widths and air
perturbation levels between 0\% and 50\% of ambient air. 
We observe that streamers in preionized and perturbed air accelerate more
efficiently than in non-ionized and uniform air with air perturbation
dominating the streamer acceleration.
We find that in unperturbed air preionization levels of $10^{11}$ m$^{-3}$
are sufficient to explain run-away electron rates measured in conjunction
with terrestrial gamma-ray flashes. In perturbed air, the production rate of runaway electrons varies
from $10^{10}$ s$^{-1}$ to $10^{17}$ s$^{-1}$ with
maximum electron energies from some hundreds of eV up to some hundreds of
keV in fields above and below the breakdown strength with only a marginal effect of the channel radius.
In the presented simulations the number of runaway
electrons matches with the number of energetic electrons measured in alignment with the observations of terrestrial
gamma-ray flashes.
Conclusively, 
the complexity of the streamer zone ahead of leader tips allows explaining
the emission of energetic electrons and photons from streamer discharges in
fields below and above the breakdown magnitudes.
\end {abstract}

\section {Introduction} \label {intro.sec}
In 1994 the Burst And Transient Source Experiment (BATSE) on the Compton
Gamma Ray Observatory (CGRO) was the first to measure beams of high-energy photons emitted from thunderstorms
\cite{fishman_1994}. These terrestrial gamma-ray flashes (TGFs) are bursts
of X- and $\gamma$-rays with photon energies ranging from several eV up to at least 40
MeV \cite{marisaldi_2010} lasting from hundreds of microseconds
\cite{briggs_2010}) up to minutes \cite{tsuchiya_2007}.
Their existence and properties have been confirmed and refined by later missions (see
e.g. \cite{smith_2005,briggs_2010,tavani_2011,tsuchiya_2011}) and are subject to the
contemporary ASIM (Atmosphere-Space Interactions
Monitor) \cite{neubert_2018} and the upcoming TARANIS (Tool for the
Analysis of RAdiation from lightNIng and Sprites) mission
\cite{blanc_2007} with payloads dedicated to the measurement of optical and
high-energy radiation emitted from thunderstorms.

Whereas it is known that these photons are Bremsstrahlung photons from
energetic electrons (e.g. \cite{torii_2004,koehn_2014a} and citations
therein), so-called runaway electrons \cite{eddington_1926,gurevich_1961}, it has not been fully understood yet how
electrons are accelerated into the energy range where they are capable of
producing photons from keV to tens of MeV. Whilst electrons are energized by the
thunderstorm electric fields, they collide in turn with air molecules and lose energies
due to inelastic collisions. Hence, there is an interplay between the electron
acceleration and the deceleration determining the characteristic electron energy
distribution function. 

The generation of runaway electrons is a stochastic process. However, its essence and magnitudes
can be explained in terms of a
conventional deterministic approach considering the simple case of a
homogeneous electric field $E$
\cite{gurevich_1961,babich_1973,kunhardt_1986,kunhardt_1988,babich_1990,babich_1995,babich_2003,babich_2005a}.
While electrons with energy $E_{kin}$ move in a dense gas medium, they experience
a drag or friction force $F(E_{kin})$ as a result of inelastic (ionization, excitation, radiative
losses) interactions with air molecules. In this deterministic approach,
a drag force is introduced as a continuous function of the electron energy
$E_{kin}$, for which either the Bethe equation
\cite{gurevich_1961,babich_2003} or more accurate
semi-empirical equations
\cite{babich_1973,babich_1990,babich_1995,babich_2005a,babich_2003} are used. Using
such continuous functions below approximately 100 eV, instead of stepwise energy losses,
is not correct since the
lost energy is comparable to the energy before the interaction.
The friction force has one maximum and one minimum, which are equal to
$F_{max}\approx 27$ MeV m$^{-1}$ at $E_{kin,max}\approx 150$ eV and
$F_{min}\approx 218$ keV m$^{-1}$ at $E_{kin,min}\approx 1$ MeV in air at
standard temperature and pressure. Above $E_{kin,min}$, the function
$F(E_{kin})$ slowly increases up to ultrarelativistic energies where radiative losses dominate.
Hence, in a homogeneous electric field with a moderate strength $E<F_{max}
e_0^{-1}$, where $e_0\approx 1.602\cdot 10^{-19}$ C is the elementary
charge, the equation $F(E_{kin})=e_0 E$ has three roots:
$E_{kin,1},E_{kin,2},E_{kin,3}$, of which $E_{kin,1}$ and $E_{kin,3}$ correspond to stable state of
the electron ensemble, whereas $E_{kin,2}$ corresponds to unstable states
\cite{babich_1995,babich_2003}. Runaway electrons are those electrons which surpass the
threshold $E_{kin,2}$ identified as the runaway threshold which is a function of
$E$ \cite{babich_1973,babich_1990,babich_1995,babich_2003,babich_2005a}. Including angular scattering increases
the friction $F_(E_{kin})$ such that $F_{max}$ and $F_{min}$ are increased
by factors of 1.5 \cite{kunhardt_1986,babich_1990,babich_2003} and of 1.25
\cite{khaerdinov_2016}. 
In extremely strong fields $E>F_{max} e_0^{-1}$, electrons are capable of
energizing up to energy $E_{kin,3}$ which, in this case, is the only root of the
equation $F(E_{kin})=e_0 E$.




There are currently two possible theories explaining the production of
high-energy run-away electrons in kilometer long lightning discharges and thunderclouds: the continuous
acceleration and multiplication of high-energy electrons, remnants from cosmic rays, in the large-scale uniform
thundercloud electric fields \cite{wilson_1925,gurevich_1992,babich_2012,gurevich_2013} or the acceleration of
low-energy electrons in the high-field regions localized close to lightning leader tips
\cite{chanrion_2008,celestin_2011,babich_2015,koehn_2015}, both with or without the feedback of Bremsstrahlung photons and
of pair-produced positrons and electrons
\cite{dwyer_2003,babich_2005,koehn_2017a}.

The formation and propagation of lightning leaders is mediated by a
multitude of streamer channels. 
The importance of these streamers on the production of runaway
electrons is manifold: Past models have indicated that electrons might be
accelerated into the run-away regime by the high electric fields at the
streamer tips \cite{celestin_2011,babich_2015} and further be accelerated
by the electric field of the lightning leader during its stepping process.
Yet, the environment of the leader tip is very complex, and there are currently no self-consistent models that
consider the influence of the streamer zone onto the environment of the
leader tip. Furthermore, Cooray et al. \cite{cooray_2009} suggested that the electric field might
significantly be enhanced during the encounter of two streamers. This is
supported by simulations by Luque \cite{luque_2017} whereas simulations by
Ihaddadene and Celestin \cite{ihaddadene_2015} and K{\"o}hn et al. \cite {koehn_2017b} have shown that the duration of
the field enhancement is too small to contribute significantly to the production of runaway electrons.

Additionally, streamers support the propagation of lightning leaders.
Several observations have indicated the stepping pattern, a discontinuous
propagation mode, of lightning leaders \cite{hill_2011,winn_2011}. Whilst the exact mechanism of leader
stepping is still under debate, the current apprehension combines the
stepping with existence of the so-called space stem \cite{reess_1995} and the
streamer corona. After the leader motion has paused, a
dipole called space stem or space leader manifests several tens of meters away.
Subsequently, streamer coronae originate from the leader tip and from both poles
of the space stem. This enables the reconnection of the two streamer
coronas facing towards each other resulting in a leader step. Afterwards
a new conducting channel is formed with the electric potential of the
old leader transferred to the former space stem. This potential drop
releases a new ionization wave becoming manifest as a streamer propagating into the
preionized channel created from the streamer corona of the space stem
averted to the leader tip side. Experiments \cite{nijdam_2011} and simulations of streamers in uniform
preionization \cite{wormeester_2010} have shown that newly incepted streamers in the above-mentioned
scenario move in preionized channel with a decay length similar to the decay
length of the streamer. Babich et al. \cite{babich_2015} have shown that for
preionization densities between $10^{10}$ m$^{-3}$ and $10^{15}$
m$^{-3}$, the production of runaway electrons is enhanced compared to the
production of runaway electrons by streamers in non-ionized air.

Along with the acceleration of electrons at the high field tips and added to the
remnants of ions, streamers also change the spatial distribution of
ambient air and thus influence their vicinity
and the proximity of lightning leader tips. Simulations by Marode et al. \cite{marode_1979} have
shown that streamer discharges heat air and initiate a radial air flow
lowering the air density close to the streamer by up to approximately 50\%
within some tens of ns. Such air perturbations have been confirmed
by more recent simulations and experiments showing that streamer
and spark discharges perturb proximate air up to
80\% \cite{eichwald_1998,eichwald_2011,kacem_2013,liu_2014,ono_2004}. In previous work, we have
examined streamer properties and modelled the production of runaway
electrons and the emission of X-rays from streamers in perturbed air
\cite{koehn_2018a,koehn_2018b}. We have observed that the production rates
and energies of high-energy electrons and photons are significantly
increased compared to those in unperturbed air.

Whereas previous streamer simulations assume no preionization, uniform
preionization or unperturbed air, the remnants of preceding streamer
channels associated to leader stepping, such as residual ions and the perturbation of ambient air,
suggest that the vicinity of streamers, and thus also of the streamer
affected leader tip zone are highly inhomogeneous. This raises the
question how such inhomogeneities influence the emission of runaway electrons and
energetic photons.

We here take one step further into more realistic modelling accounting for
the preionization and air perturbations associated to leader stepping
and explore their relative importance for the production of
runaway electrons. Including these two effects, we determine streamer
properties as well as the fluence and maximum energies of runaway electrons. Finally, we conclude
which conditions favour the production of energetic electrons serving
as a seed for the development of secondary run-away electron avalanches
and thus also for energetic photons.

\section {Modelling} \label {model.sec}
\subsection {Set-up of the simulation domain and introduction of the Monte
Carlo model} \label{setup.sec}
\begin {figure}
\includegraphics [scale=0.46] {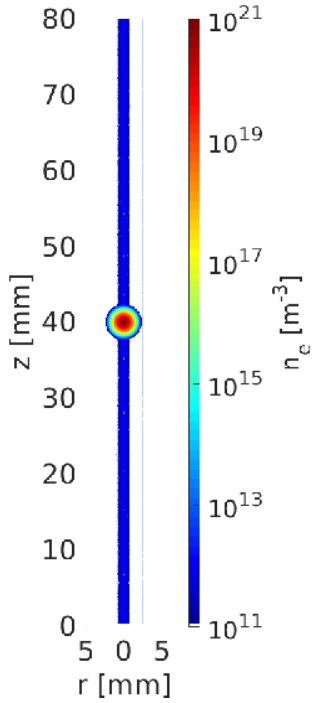}
\caption {The simulation domain showing the electron density of the initial
plasma patch (\ref{dens.1}) and of the preionization channel defined by Eq. (\ref{pre.1}) 
$n_{pre,0}=10^{12}$ m$^{-3}$. The preionized air channel
is radially extended as widely as the initial electron-ion patch which is not visible
because the colorbar is limited down at $n_{e}=10^{11}$ m$^{-3}$.} \label{domain.fig}
\end {figure}

We here employ a 2.5D cylindrical particle-in-cell Monte Carlo code with two spatial
$(r,z)$ and three velocity coordinates $(v_r,v_z,v_{\theta})$ which
has been used before (see e.g. \cite{chanrion_2008,koehn_2018a,koehn_2018b}) and allows us to trace individual (super)electrons
as well as to monitor the formation of bipolar streamers from a charge-neutral
electron-ion patch
\begin {eqnarray}
n_{e,i}(r,z,t=0)=n_{e,0}\cdot\exp\left(-(r^2+(z-z_0)^2)/\lambda_0^2\right) \label{dens.1}
\end {eqnarray}
centered in the middle of the simulation domain, i.e. $z_0=L_z/2$,
with a peak density of $n_{e,0}=10^{20}$ m$^{-3}$ and a Gaussian length of
$\lambda_0=0.5$ mm. 

The size of the simulation domain, displayed in Fig.
\ref{domain.fig}, is $(L_r,L_z)$=(6 mm, 80 mm) (as in \cite{babich_2015}) on
a mesh with $150\times 1600$ grid points.
This grid is used to solve the Poisson equationä
\begin {eqnarray}
\Delta\phi=e_0/\epsilon_0\cdot\left(n_i-n_e\right) \label{poisson.1}
\end {eqnarray}
for the electrostatic potential $\phi$ taking into account the effect of space charges. At the boundaries
$r=0,L_r$, we use the Neumann condition $\partial\phi/\partial r=0$, and atä
the boundaries $z=0,L_z$, we use the Dirichlet conditions $\phi(r,0)=0$ and
$\phi(r,L_z)=E_{amb}\cdot L_z$ where $E_{amb}$ is the ambient electric
field. We here consider two different ambient fields, $E_{amb}=50$ kVä
cm$^{-1}\approx 1.56E_k$ \cite{babich_2015} and $E_{amb}=0.5E_k$ where we
here and throughout the paper refer to $E_k\approx 3.2$ MV m$^{-1}$ as the classical breakdown field in air at
standard temperature and pressure (STP). In the current simulation set-up, the applied electric fields are
equivalent to voltages of 400 kV and 128 kV.

We here trace individual (super)electrons interacting with ambient air.
Unlike fluid models, tracing individual (super)electrons with a particle
code allows us not only to obtain streamer properties such as the electron
density or electric field distribution, but also to estimate the electron
energy distribution. We include electron impact ionization, elastic and inelastic
scattering as well as electron attachment and bremsstrahlung. Additionally, we apply
a photoionization model where photons emitted from excited nitrogen
ionize oxygen molecules locally and liberate additional electrons. More details of
the applied Monte Carlo model are described in
\cite{chanrion_2008,koehn_2017}.

Since electrons ionize molecular nitrogen and oxygen, the electron number grows
exponentially leading to an electron avalanche and eventually a streamer.
Due to limited computer memory, we use an adaptive particle
scheme \cite{chanrion_2008} conserving the charge distribution as well as
the electron momentum such that every simulated electron is a
superelectron representing $w$ physical electrons.

\subsection {Implementation of air perturbations} \label{perturb.sec}
\begin {figure}
\includegraphics [scale=0.56] {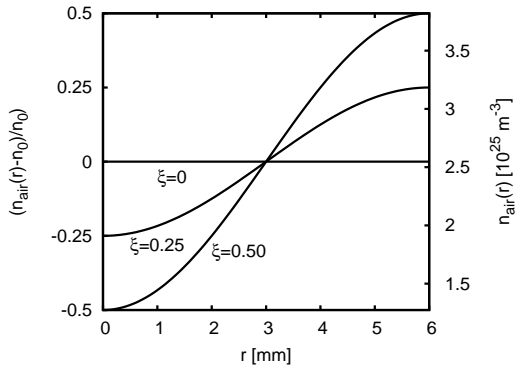}
\caption {Air density (\ref{dens.2}) as a function of $r$ for perturbations of 0,
25\% and 50\%} \label {dens.fig}
\end {figure}

In Monte Carlo particle simulations, we include the collisions of electrons with
ambient air where the nitrogen and oxygen molecules are put at random
positions as an implicit background. The probability $P_c$ of a collision of
an air molecule with an electron with velocity
$v_e$ within the time interval $\Delta t$ is
$P_c=1-\exp\left(-n_{air}v_e\sigma\Delta t\right)$ where $n_{air}$ is the number density
of ambient air and $\sigma$ the collision cross section.

Previous experiments and simulations \cite{plooster_1970,marode_1979,eichwald_1996,ono_2004,kacem_2013} suggest that
shock waves and thermal expansion by leaders and also by the small-scale
discharge modes, the streamers, are capable of perturbing the vicinity of their location
up to 80\% of the ambient air level \cite{kacem_2013}. For $n_{air}$,
we therefore choose the ansatz
\begin {eqnarray}
n_{air}(r)=n_0\left(1-\xi\cos\left(r\cdot\pi/L_r\right)\right) \label{dens.2}
\end {eqnarray}
with a global minimum on the symmetry axis ($r=0$) and a global maximum on the outer
boundary ($r=L_r$). The sinusoidal form has been computed by Marode et al.
\cite {marode_1979} and is here meant to capture the minimum air density in
the proximity of the channel axis driving air molecules to the exterior
boundary. Otherwise, the actual form of $n_{air}$ is not
crucial and we here limit ourselves to $\xi=0,0.25,0.5$ neglecting very large
perturbations in the vicinity of the streamer channel. Note that the time $t_D$ of air molecules to diffuse back to uniform density is in
the order of $t_D\simeq L_r^2/D_{air}\approx$ 1.8 s with $D_{air}\approx 2\cdot 10^{-5}$
m$^2$ s$^{-1}$ \cite{cussler_1997} which is much larger than the simulation time
of the order of several nanoseconds allowing us to assume a stationary
distribution of air molecules.

\subsection {Implementation of preionization} \label{preioniz.sec}
As discussed by Babich et al. \cite{nijdam_2011,babich_2015}, streamers leave behind
residual ionization affecting the motion of successive streamer and
leader channels. The reminiscent density $n_{pre}$ of the previous streamer channel
is modelled by 
\begin {eqnarray}
n_{pre}=n_{pre,0}\cdot\exp(-r^2/\lambda_{pre}^2) \label{pre.1}
\end {eqnarray}
where $n_{pre,0}=10^{10}-10^{13}$ m$^{-3}$ determines the peak density and
$\lambda_{pre}=0.5\lambda_0,1.0\lambda_0,1.5\lambda_0$ the width of the preionized channel.
This approach is advocated, firstly because each streamer discharge
has its own characteristic minimal radius depending on the streamer velocity
and the ambient gas density \cite{naidis_2009}, secondly because
the charge and the width of the preionized channel diffuse with time \cite{nijdam_2011}.

After a preceding discharge, the time to readjust the electric field is in
the order of some ns-$\mu$s \cite{koehn_2018b} which is significantly smaller than
the diffusion time $t_D$. Hence, the screening of the electric field is negligible in the
current set-up which justifies to run simulations in $E_{amb}=1.56E_k$.

Fig. \ref{domain.fig} shows the initial electron-ion patch together
with the preionized channel ($\lambda_{pre}=\lambda_0$ and n$_{pre,0}=10^{12}$
m$^{-3}$). It illustrates how the initial electron-ion patch is embedded in the
preionized channel. Note that the channel is extended radially as much as
the electron-ion patch which is not visible because the colorbar is
limited down at $n_e=10^{11}$ m$^{-3}$.

\section {Results} \label {results.sec}
\subsection {Benchmarking} \label {bench.sec}
\begin {figure}
\begin {center}
\begin {tabular}{cc}
\includegraphics [scale=0.56] {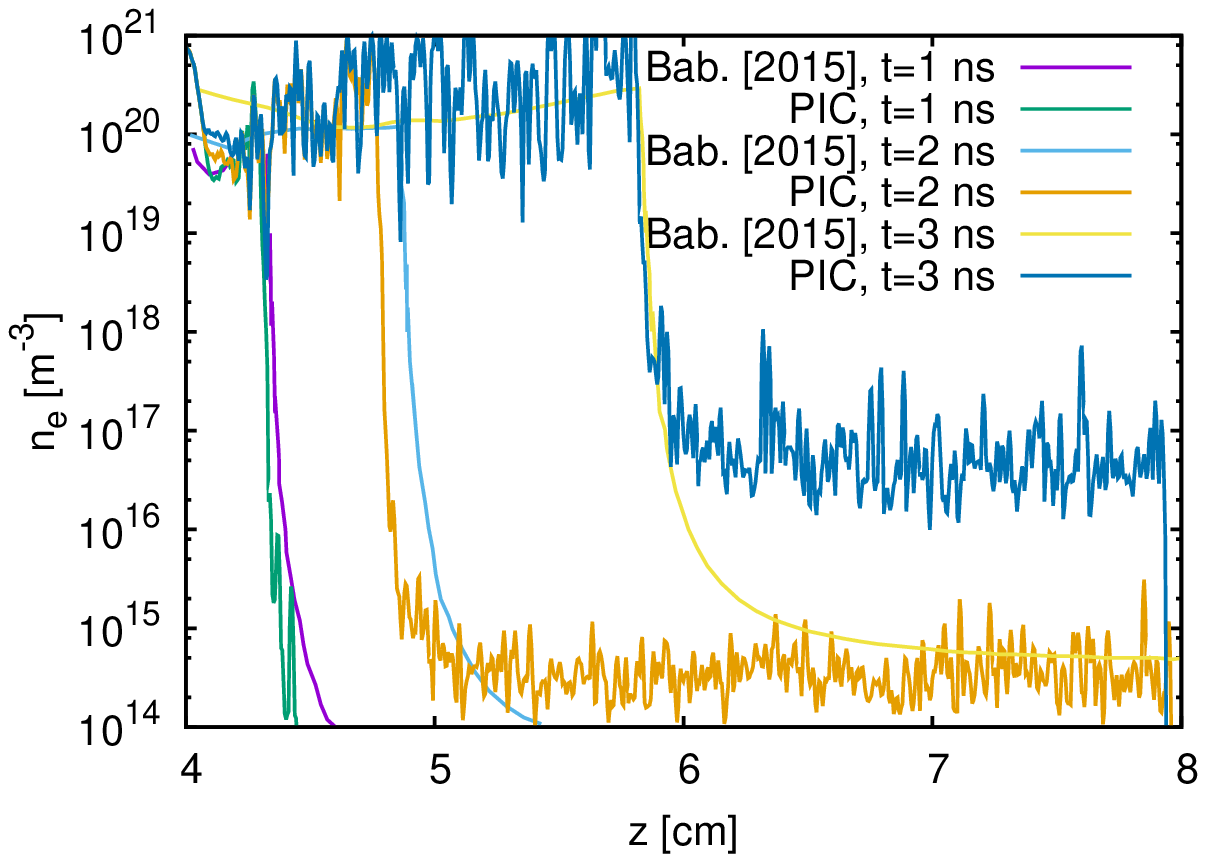} &
\includegraphics [scale=0.56] {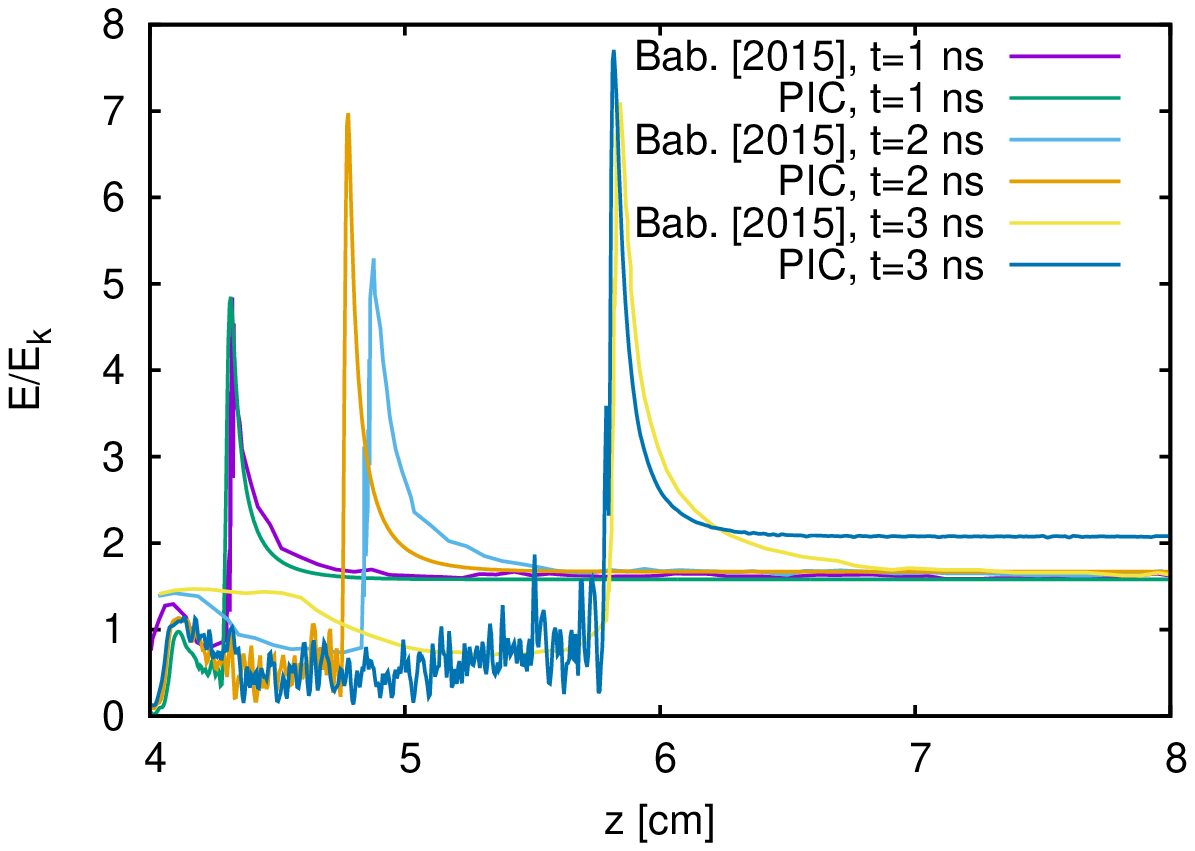} \\
a) $n_e$ & b) $E$
\end {tabular}
\caption {The on-axis electron density (a) and the on-axis electric field
(b) as a function of $z$ calculated from the simulations by Babich et al.
\cite{babich_2015} and from particle-in-cell (PIC) simulations for a preionization level of
$10^{12}$ m$^{-3}$ without any air perturbation.} \label {bench.fig}
\end {center}
\end {figure}

Babich et al. \cite{babich_2015} have already solved the fluid equations of
negative streamers in preionized air without the effect of air perturbations focusing on the
production of run-away electrons in
a field of $E_{amb}=50$ kV cm$^{-1}$. Fig. 
\ref{bench.fig} compares the on-axis electron density (a) and the on-axis
electric field (b) of the negative streamer front for $n_{pre,0}=10^{12}$ m$^{-3}$ computed by Babich et al.
and computed by MC particle simulations showing a very good agreement in the streamer
channel. There is a slight deviation after 2 ns which
is compensated again after 3 ns. In all considered cases, however, our results show
fluctuations which do not occur in the previous results.
Yet, this is not surprising since a particle code normally shows more fluctuations than a fluid code, see. e.g. \cite{li_2012}. At the tips, the electric
field peaks as smoothly as for the fluid code. Beyond the streamer channel
the electron density is larger than the
electron density calculated by the fluid equations. This discrepancy, however, is
not relevant since we are interested in the properties of the main streamer
channel and since the production of runaway electrons predominantly takes place in
the streamer head and not beyond.

\subsection {Streamer evolution in uniform, preionized air} \label {pre_1.sec}
\begin {figure}
\begin {center}
\begin {tabular}{ccc}
\includegraphics [scale=0.45] {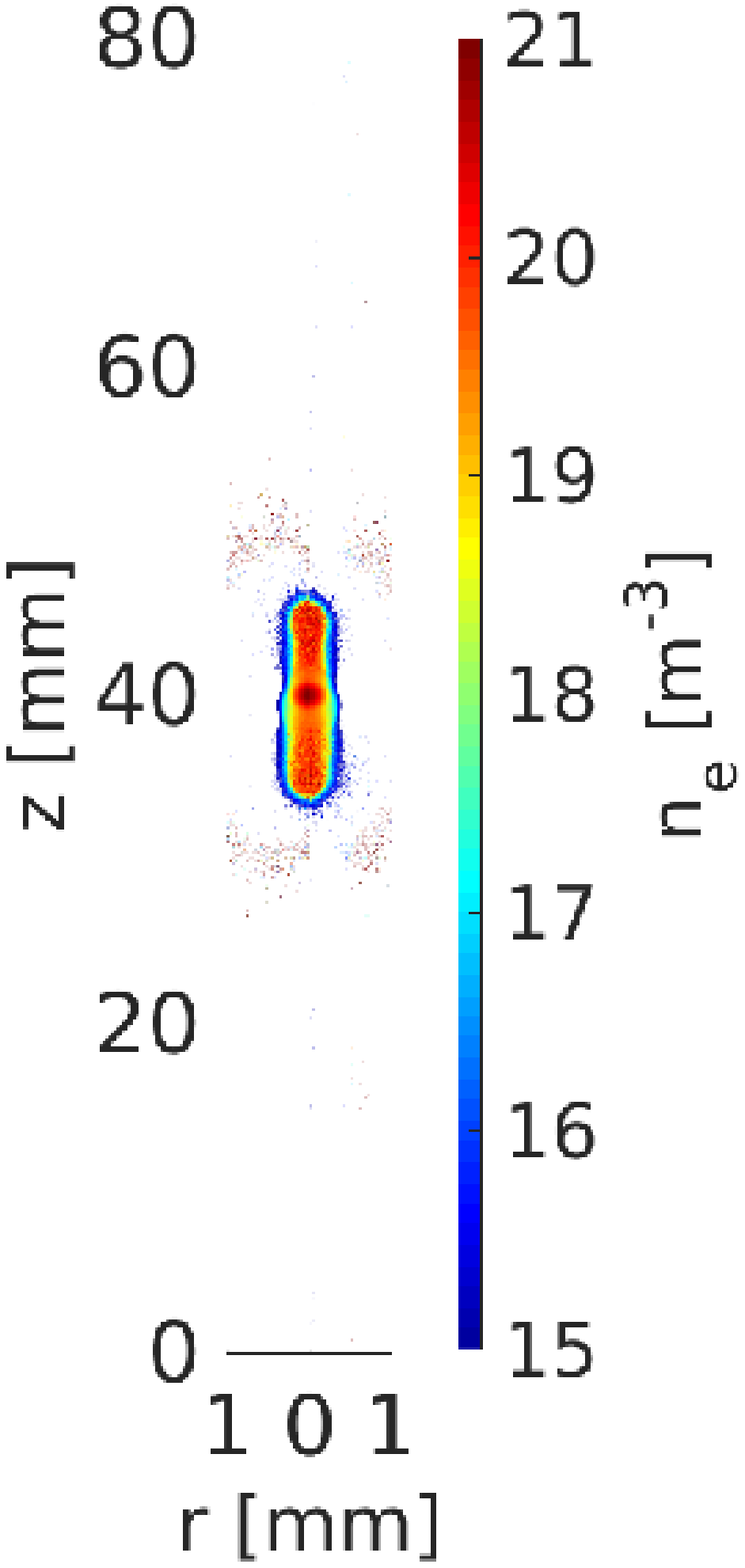} &
\includegraphics [scale=0.45] {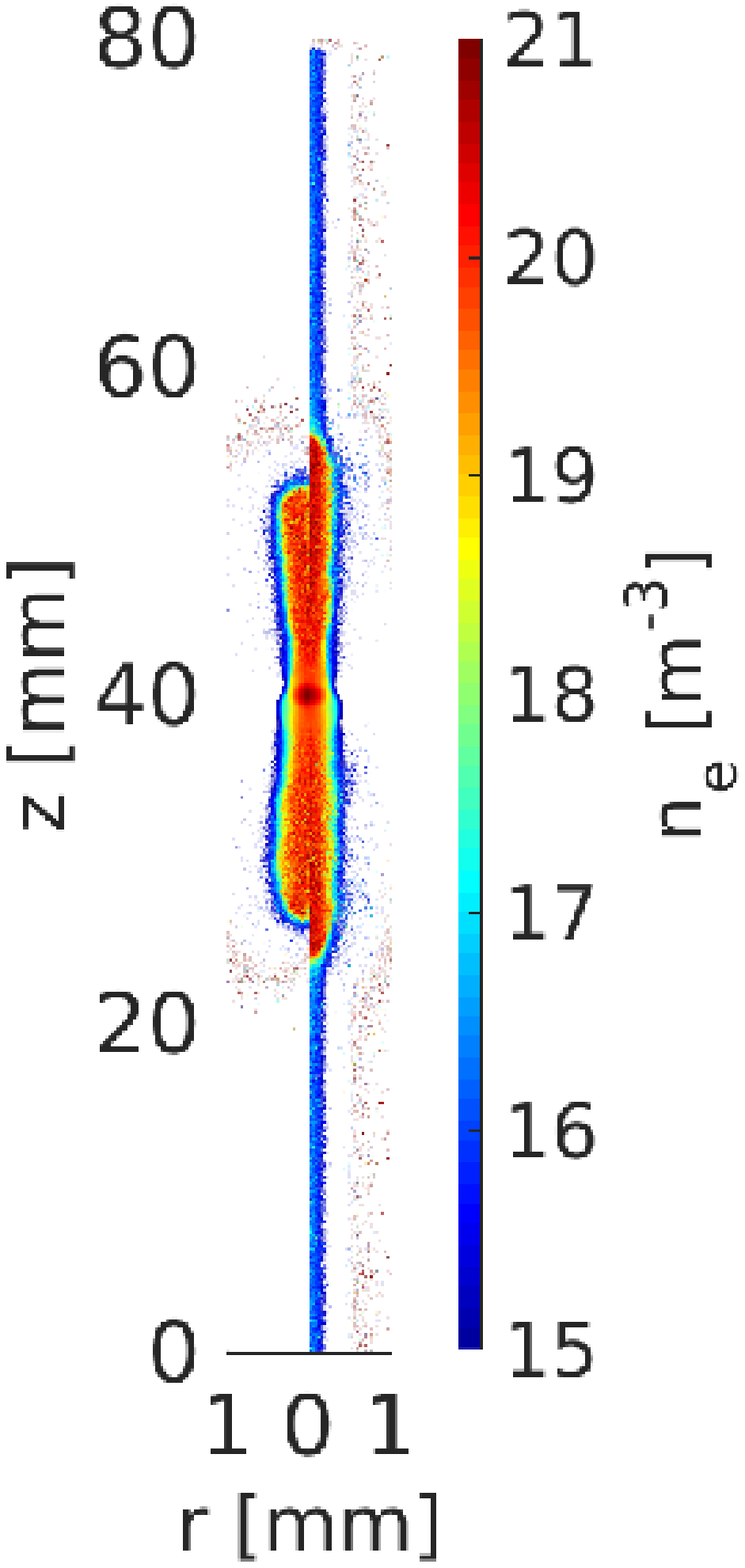} &
\includegraphics [scale=0.45] {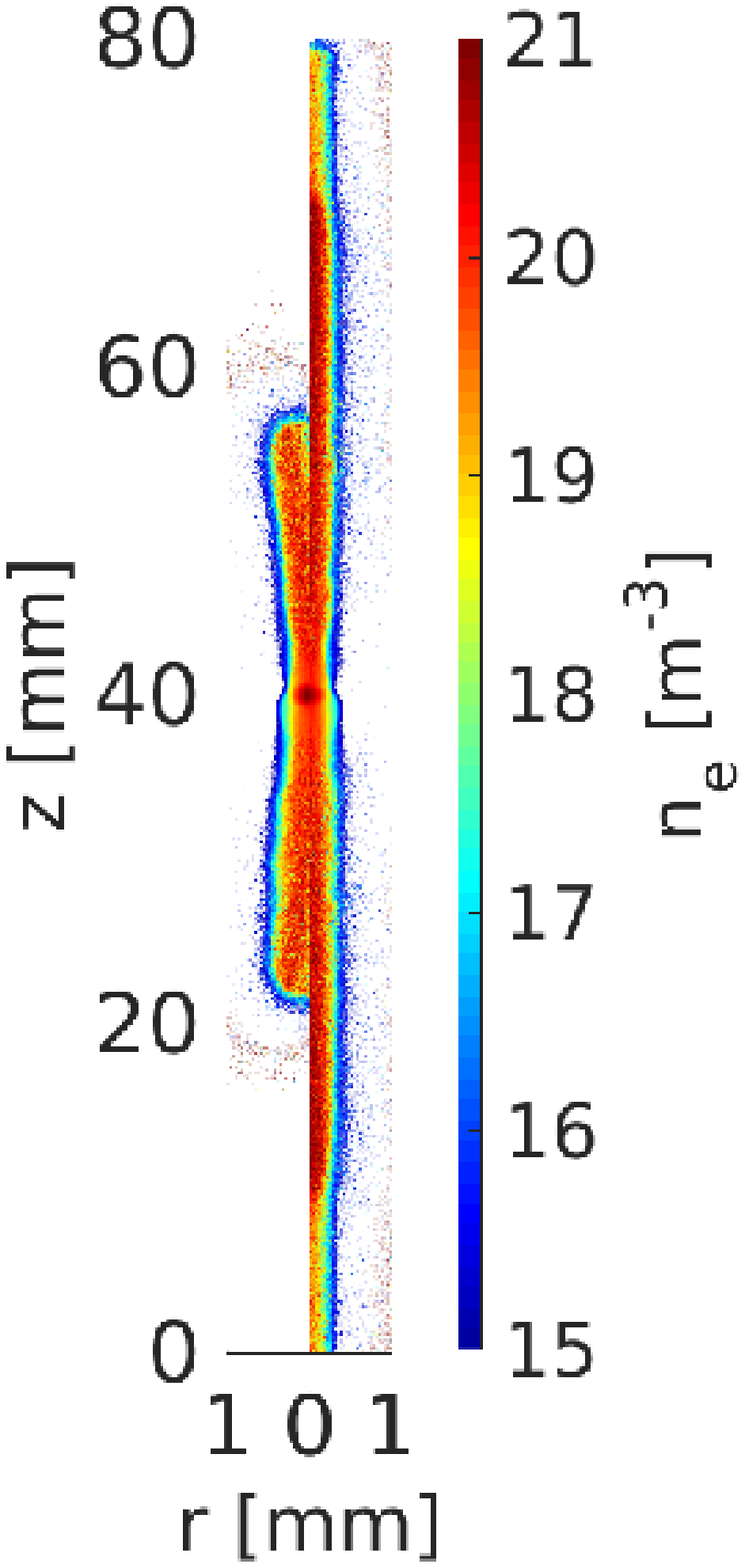} \\
\includegraphics [scale=0.45] {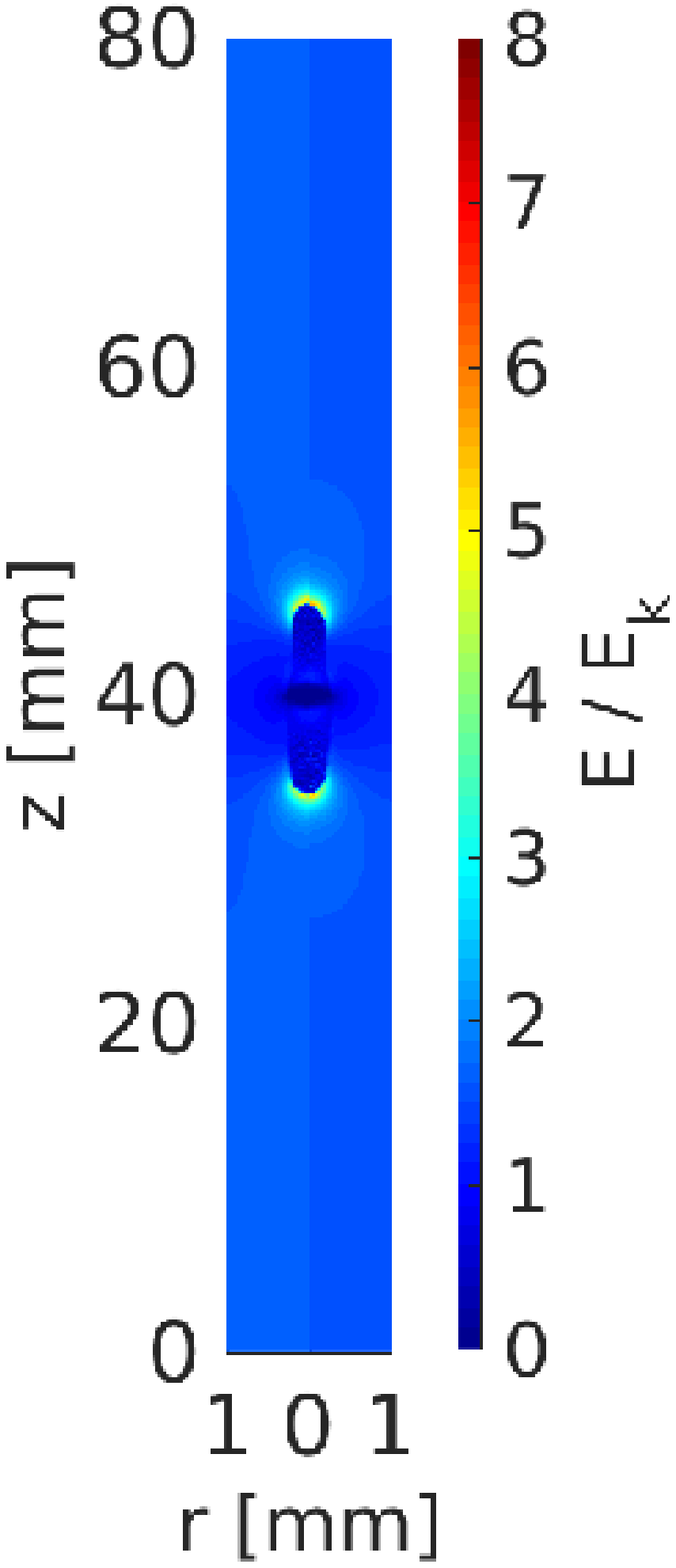} &
\includegraphics [scale=0.45] {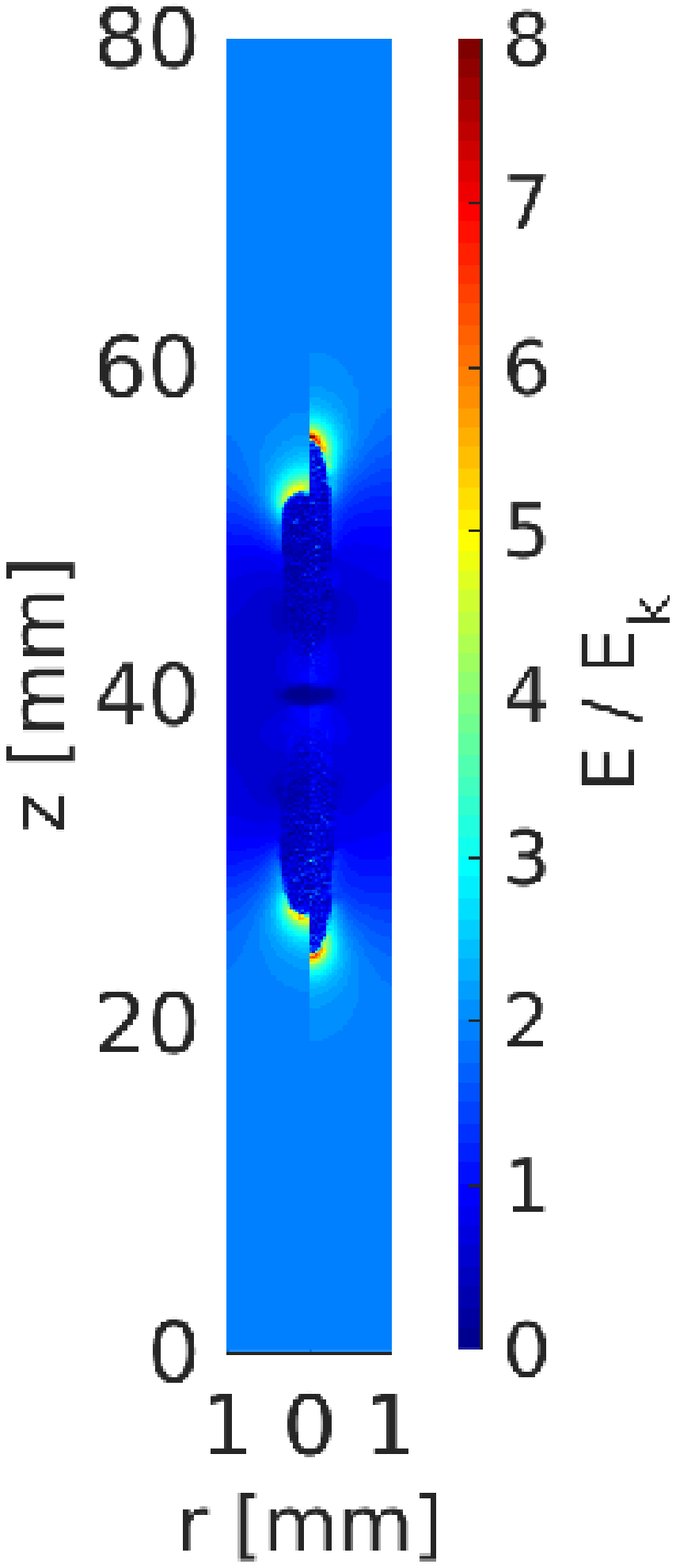} &
\includegraphics [scale=0.45] {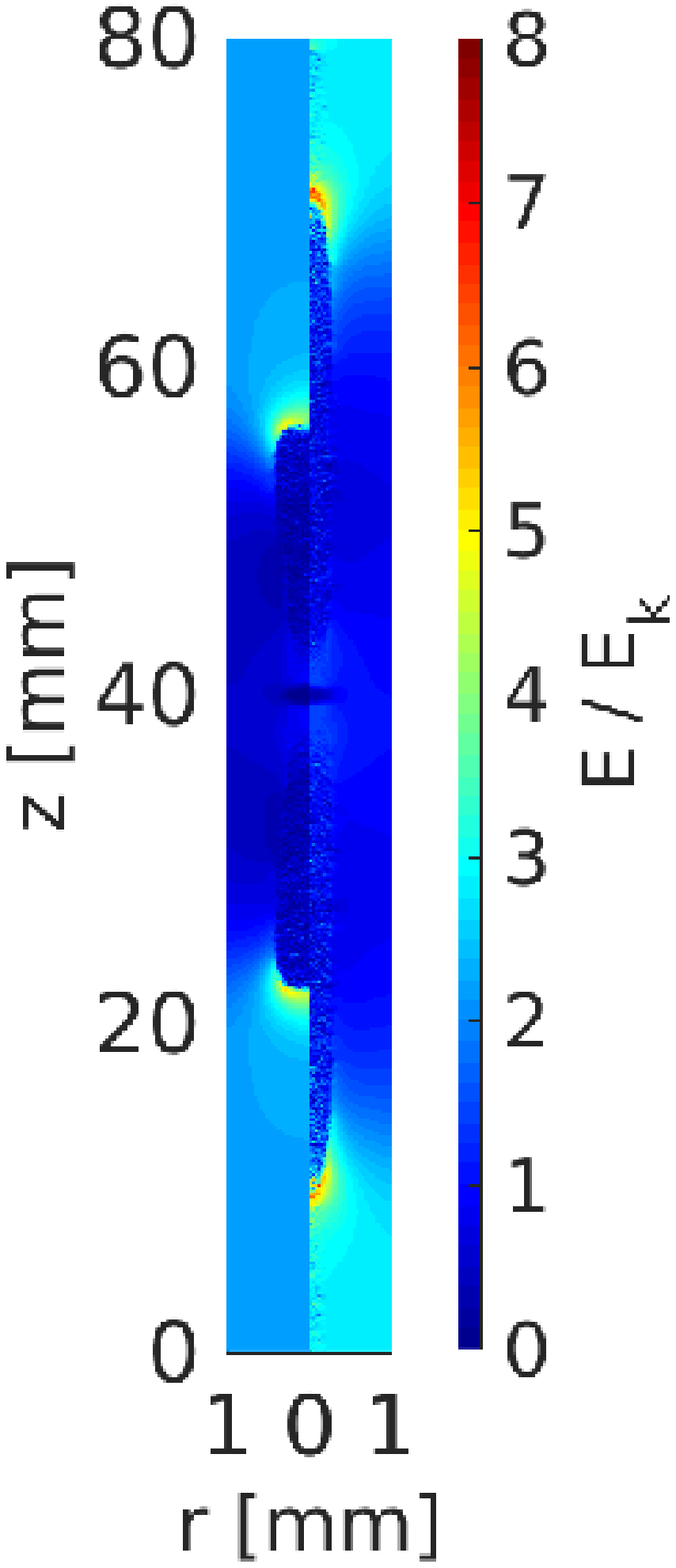} \\
$t=1.62$ ns & $t=2.84$ ns & $t=3.44$ ns
\end {tabular}
\end {center}
\caption {The electron density (top) and the electric field (bottom)
in non-ionized air (left half of each panel) and in preionized air with
$n_{pre,0}=10^{12}$ m$^{-3}$ (right half) after different time steps.} \label {dens_1.fig}
\end {figure}

\begin {figure}
\begin {center}
\begin {tabular}{cc}
\includegraphics [scale=0.56] {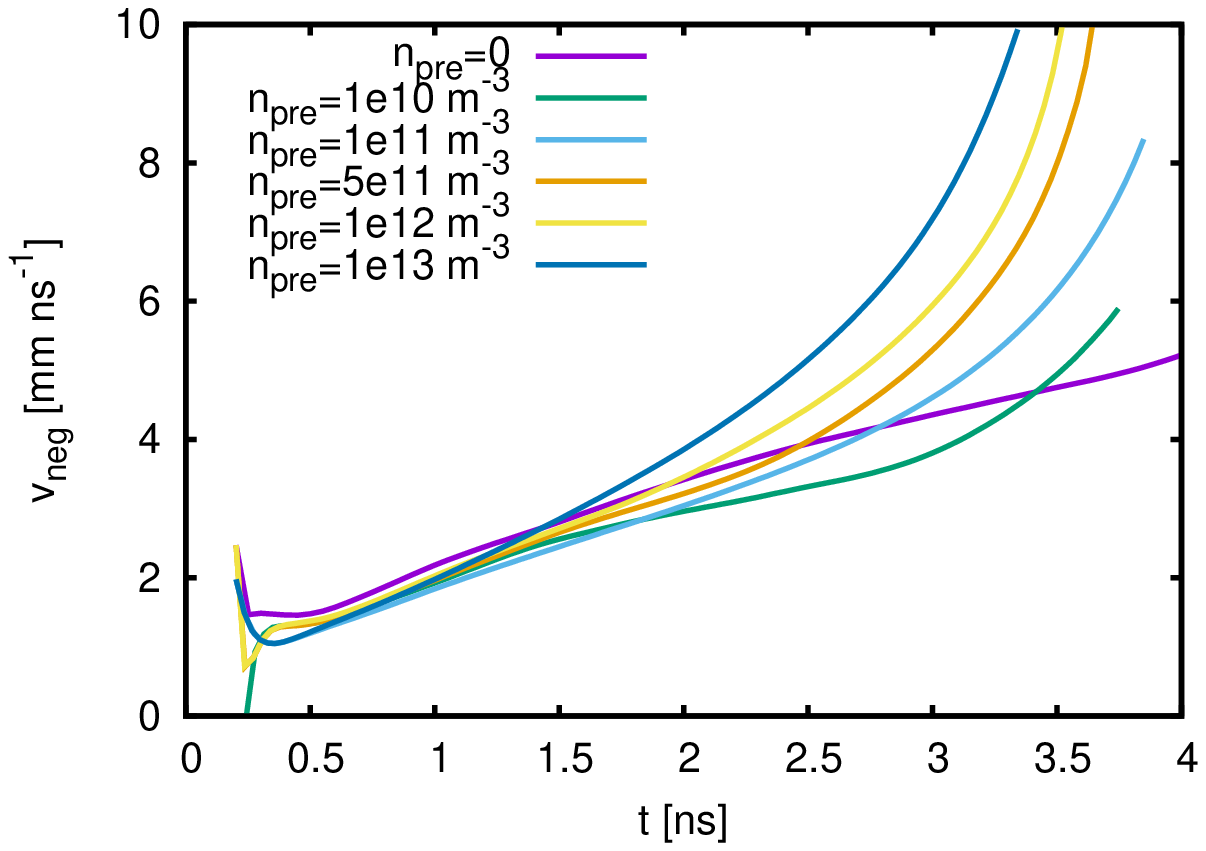} &
\includegraphics [scale=0.56] {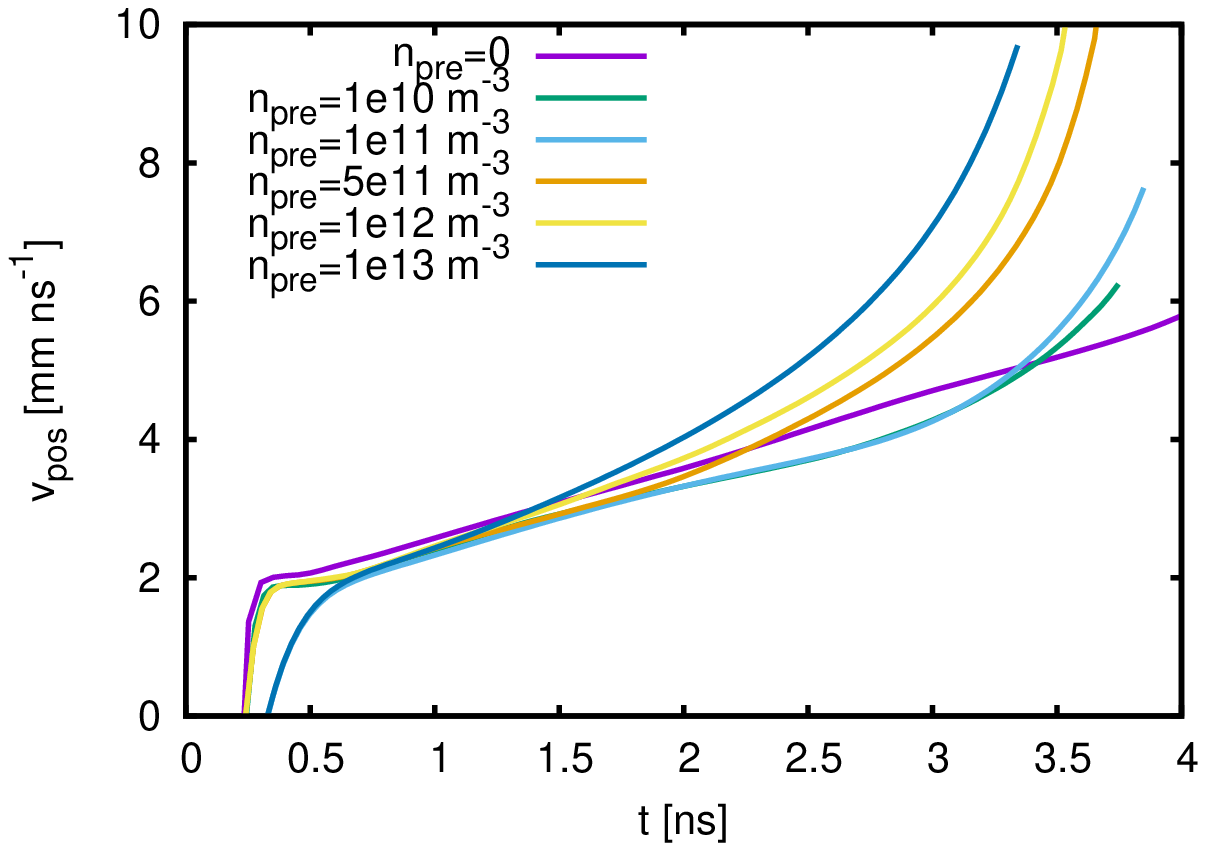} \\
a) $v_{neg}(t)$ & b) $v_{pos}(t)$\\
\includegraphics [scale=0.56] {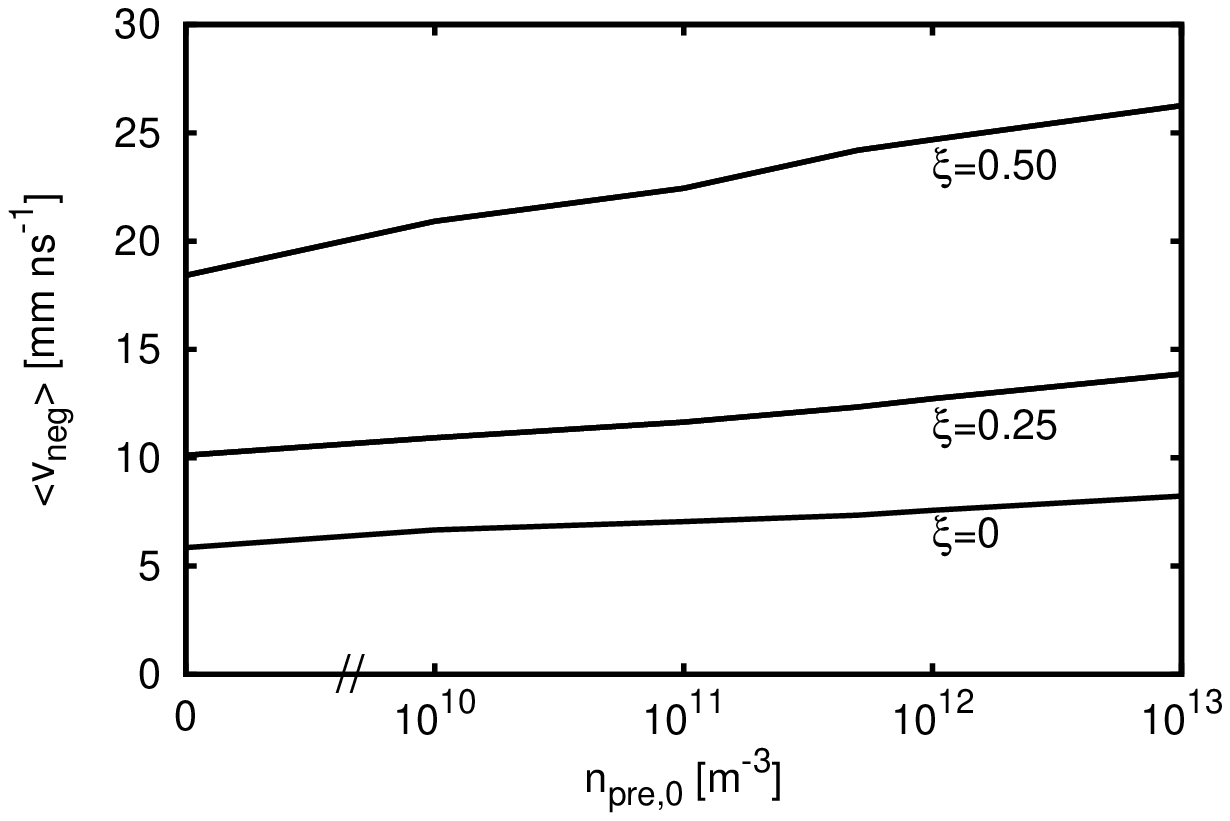} &
\includegraphics [scale=0.56] {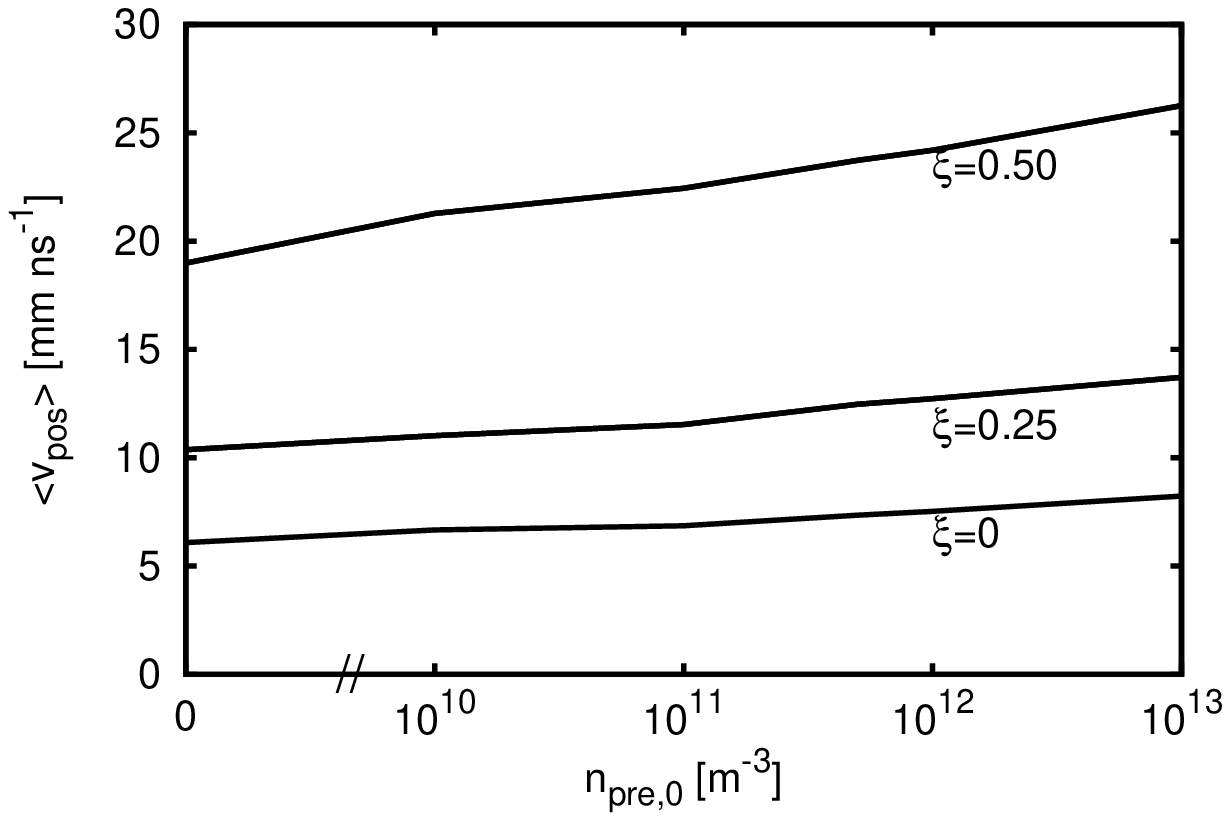} \\
c) $\langle v_{neg} \rangle$ & d) $\langle v_{pos} \rangle$
\end {tabular}
\end {center}
\caption {The velocities of the negative (a) and positive (b) streamer front in
uniform air for different levels of preionization as a function of time.
The mean front velocities of the negative (c) and positive (d) streamer front.} \label {veloc_1.fig}
\end {figure}

We here commence our study of the streamer evolution in preionized air only,
similar to the set-up discussed in \cite{babich_2015}.

Figure \ref{dens_1.fig} compares the electron density and the electric field
in non-ionized air with the electron density and the electric field in
preionized air with $n_{pre,0}=10^{12}$ m$^{-3}$. After 1.62 ns the streamer
length and the electric field are equally large irrespective of $n_{pre,0}$.
After 2.84 ns, however, the streamer in preionized air overtakes the streamer in non-ionized air and the field
at the streamer tips is slightly more enhanced. In addition
the density of the preionized channel has reached values of above $10^{15}$
m$^{-3}$ distributed uniformly beyond the streamer tips which is not the
case in the absence of preionization. Finally,
after 3.44 ns, the streamer channel in preionized air has completely grown
into the preionized channel (within the simulation domain) and has proceeded approximately twice as much as
the streamer in non-ionized air.

Figure \ref{veloc_1.fig} a) and b) show the streamer velocities of the negative and
positive fronts in non-perturbed air. It shows that initially streamers move comparably
fast for different levels of preionization. After a few ns, however, streamers
in the highest level of preionization, here $n_{pre,0}=10^{13}$ m$^{-3}$, accelerate more effectively than
streamers in non-ionized air, followed by streamers in air with descending
order of preionization.

The different acceleration of streamers in non-ionized and in
preionized air is an artefact of the space-charge induced
electric field. In the early stages of the streamer development there is no significant
contribution of the preionized air channel. However, after several time steps
depending on $n_{pre,0}$, the streamers grow into the preionized channel
and hence the electric field at the tips energizes
the channel electrons in the vicinity of the streamer head. These channel
electrons subsequently gain enough energy to ionize molecular nitrogen and
oxygen and create additional
space charge in the proximity of the streamer head; thus, the electric
field and the velocities of streamers in highly ionized air exceed the field and velocities of
streamers in less ionized air.

\subsection {Run-away electron production in uniform and preionized air} \label {pre_2.sec}
\begin {figure}
\includegraphics [scale=0.56] {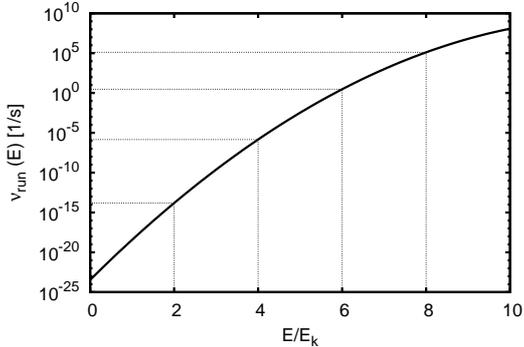}
\caption {The runaway rate $\nu_{run}$ (\ref{runaway_1}) as a function of
the electric field.} \label{runaway_rate.fig}
\end {figure}

\begin {figure}
\includegraphics [scale=0.56] {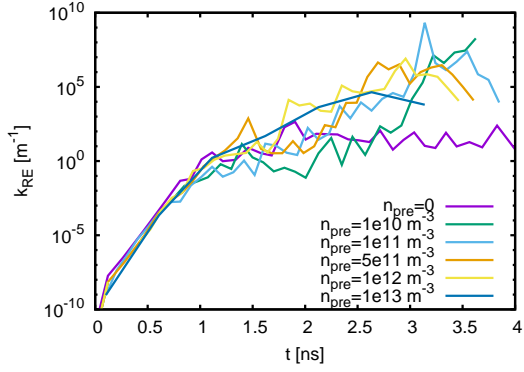}
\caption {The number $k_{RE}(t)=dN_{RE}/dz$ of produced runaway electrons per unit
length (\ref{runaway_2}) as a function of time for different levels of preionization in
uniform air.} \label {runaway_1.fig}
\end {figure}

\begin {figure}
\begin {center}
\begin {tabular}{cc}
\includegraphics [scale=0.56] {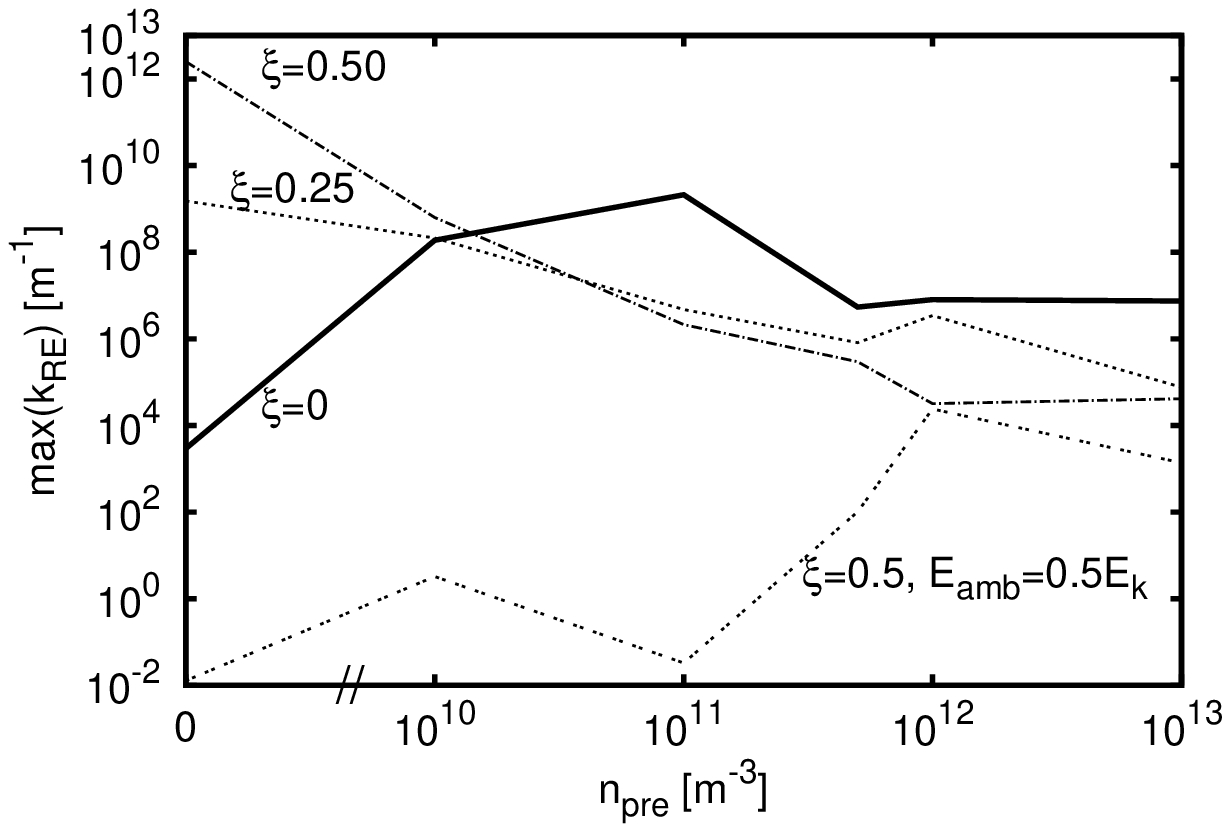} &
\includegraphics [scale=0.56] {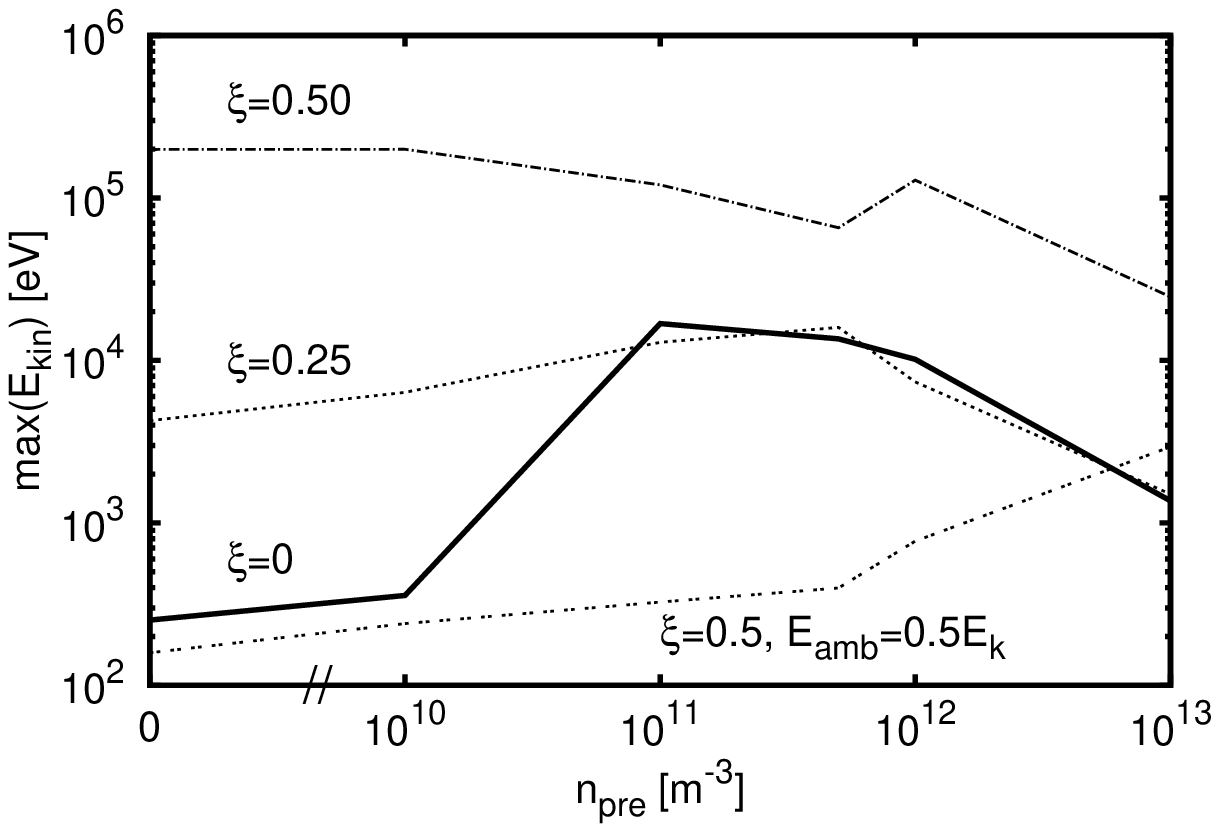}\\
a) $\max\limits_{t}\left(k_{RE}\right)$ for $\lambda_{pre}=\lambda_0$ & b) $\max\limits_{t}\left(E_{kin}\right)$ for $\lambda_{pre}=\lambda_0$  \\
\includegraphics [scale=0.56] {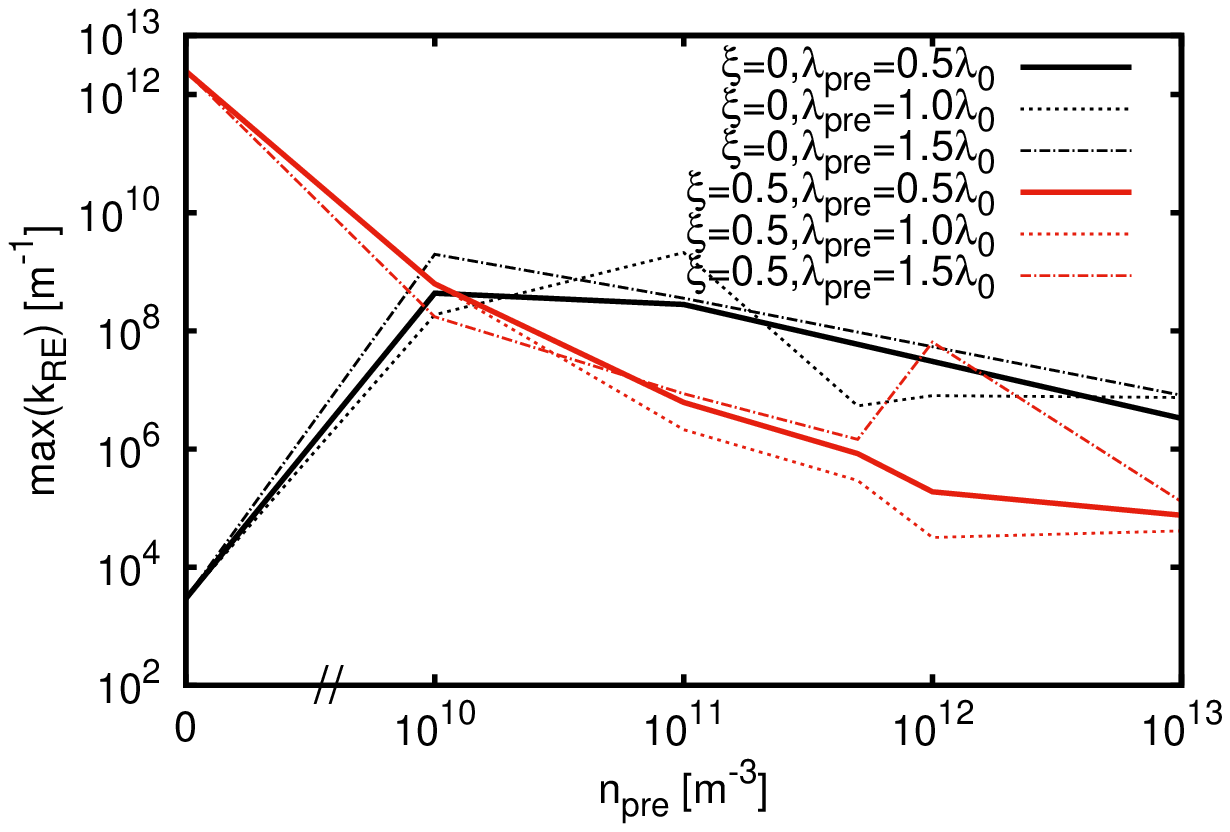} &
\includegraphics [scale=0.56] {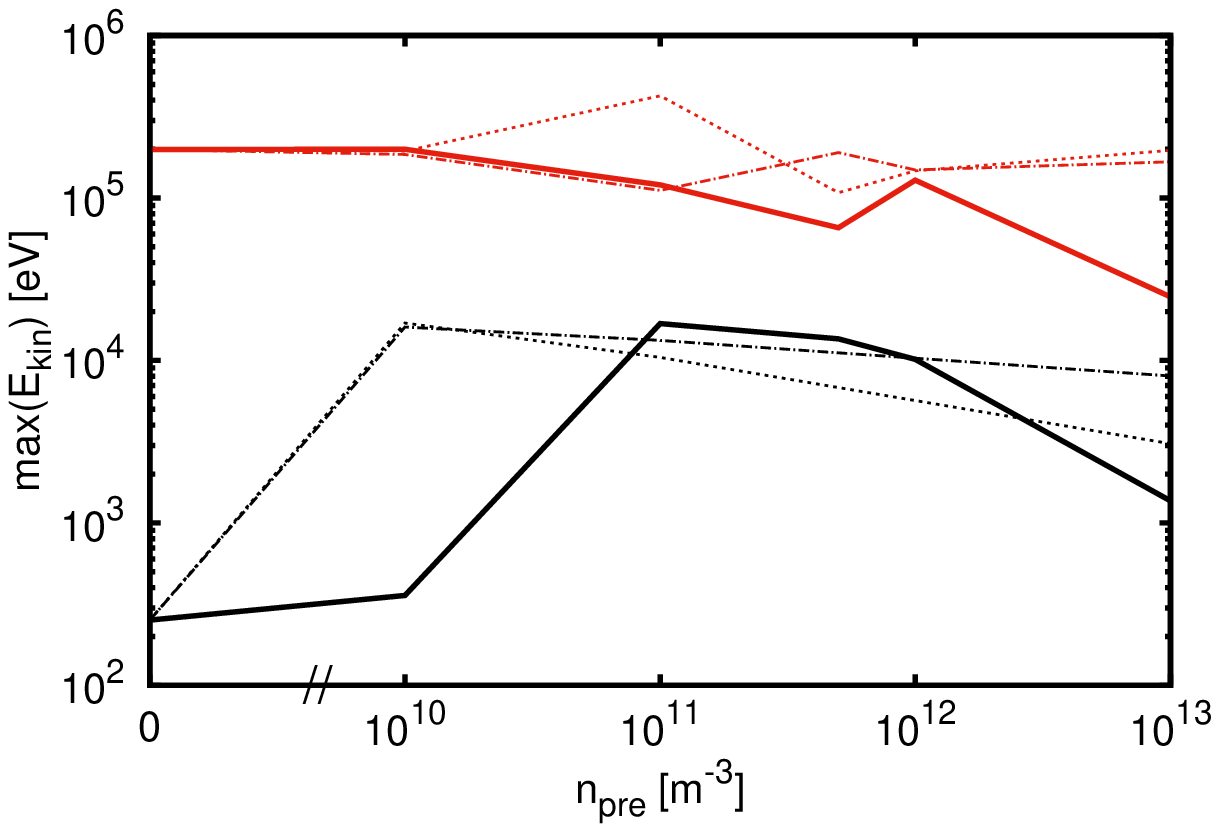}\\
c) $\max\limits_{t}\left(k_{RE}\right)$ for different $\lambda_{pre}$ & d) $\max\limits_{t}\left(E_{kin}\right)$ for different $\lambda_{pre}$
\end {tabular}
\end {center}
\caption {The maximum number of runaway electrons per unit length,
$\max\limits_{t}\left(k_{RE}\right)$, as a function of $n_{pre,0}$ for
$\lambda_{pre}=\lambda_0$ (a) and for different $\lambda_{pre}$ (c). The maximum electron
energy $\max\limits_{t}\left(E_{kin}\right)$ for $\lambda_{pre}=\lambda_0$
(b )and for different $\lambda_{pre}$ (d). If not denoted otherwise, the
ambient field here amounts to $1.56E_k$.} \label {runaway_2.fig}
\end {figure}

\begin {table}
\begin{tabular}{c|c|c}
$n_{pre,0}$ [m$^{-3}$] & $\max\limits_t k_{RE}(t)$ [m$^{-1}$] & $k_{RE,Babich}$ [m$^{-1}$] \\
\hline
$0$ & $2.92\cdot 10^3$ & --------\\
$10^{10}$ & $1.88\cdot 10^8$ & $2\cdot 10^4$\\
$10^{11}$ & $2.12\cdot 10^9$ & $5\cdot 10^7$\\
$5\cdot 10^{11}$ & $1.68\cdot 10^6$ & --------\\
$10^{12}$ & $8.06\cdot 10^6$ & $2\cdot 10^9$\\
$10^{13}$ & $7.48\cdot 10^6$ & $9\cdot 10^8$
\end {tabular}
\caption {The maximum number of produced runaway electrons per unit
length, $\max\limits_{t}\left(k_{RE}\right)$, as a function of $n_{pre,0}$.
For comparison, we also show $k_{RE,Babich}$ (\ref{runaway_5}).} \label {runaway.tab}
\end {table}

Let us now turn to the production properties of runaway electrons from
streamers in preionized air as they might occur after the reconnection of
the space stem with a stepping leader.

The runaway rate $\nu_{run}(E)$ at standard temperature and pressure as a function of the electric field
strength, i.e. the number of runaway electrons per unit time is \cite{bakhov_2000}
\begin {eqnarray}
\nu_{run}(E)&=&3.5\cdot 10^{-24}\ \textnormal{s}^{-1}\nonumber\\
&\times& \exp\left(-
\left(2.166\cdot 10^{-7}\ \textnormal {m V}^{-1}\cdot E\right)^2+3.77\cdot
10^{-6}\ \textnormal{m V}^{-1}\cdot E\right). \nonumber\\      \label{runaway_1}
\end {eqnarray}
Figure \ref{runaway_rate.fig} shows that for fields of up to $10E_k$ $\nu_{run}$
varies between approximately $10^{-24}$
s$^{-1}$ and $10^8$ s$^{-1}$. For electric fields above
$6E_k$, it is $\nu_{run}(E=6E_k)\approx 2.94$ s$^{-1}$,
$\nu_{run}(E=7E_k)\approx 990.66$ s$^{-1}$ and $\nu_{run}(E=8E_k)\approx
1.28\cdot 10^5$ s$^{-1}$; hence, the local electric field has a significant
effect on the production rate of runaway electrons. The rate $\nu_{run}$
allows us to estimate the number $k_{RE}$ of runaway
electrons per unit length
\begin {eqnarray}
k_{RE}(t)=\frac{d N_{RE}}{dz}=\frac{\frac{dN_{RE}(t)}{dt}}{\frac{dz(t)}{dt}}
=\frac{\dot{N}_{RE}(t)}{v(t)} \label {runaway_2}
\end {eqnarray}
where $v$ is the streamer velocity and where the number $N_{RE}$ of runaway electrons is \cite{babich_2015}
\begin {eqnarray}
N_{RE}(t)&=&\int\limits_{\bar{t}=0}^t \int\limits_{\bar{V}_{sim}} \nu_{run}\left(E\left(\bar{r},\bar{z},\bar{t}
\right)\right)\cdot n_e\left(\bar{r},\bar{z},\bar{t}\right) d\bar{V}
d\bar{t}   \label {runaway_3} \\
&=& 2\pi\int\limits_{\bar{t}=0}^t \int\limits_{\bar{r}=0}^{L_r}
\int\limits_{\bar{z}=0}^{L_z} \nu_{run}\left(E\left(\bar{r},\bar{z},\bar{t}
\right)\right)\cdot n_e\left(\bar{r},\bar{z},\bar{t}\right) \bar{r} d\bar{r}
d\bar{z} d\bar{t}. \label {runaway_4}
\end {eqnarray}

Figure \ref{runaway_1.fig} shows the number $k_{RE}(t)$ of runaway electrons per unit
length in non-perturbed air as a function of time.
Within approximately the first ns, there is no apparent effect of
preionization. After 1-2 ns, $k_{RE}(t)$ increases differently fast
depending on the preionization level. $k_{RE}(t)$ grows largest for
$n_{pre,0}=10^{11}$ m$^{-3}$ whereas levels of
preionization below and above $10^{11}$ m$^{-3}$ have a less dominant effect
on the production of energetic electrons. Table \ref{runaway.tab} and Fig.
\ref{runaway_2.fig} a) show that
$\max\limits_{t}\left(k_{RE}\right)$, the maximum of $k_{RE}(t)$ over time, is
smallest in non-ionized air and increases until $n_{pre,0}=10^{11}$ m$^{-3}$. This increase results from
the more enhanced electric field at the streamer tips for larger preionization.
In turn, above $n_{pre,0}=10^{11}$ m$^{-3}$, the high electron density in
the vicinity of the streamer front screens
the electric field at the tips and thus limits the maximum field strength
and the maximum number of runaway electrons.

Table \ref{runaway.tab} also compares our results with the runaway rate
$k_{RE,Babich}$ by Babich et al. \cite{babich_2015} which is defined in a
slightly different manner:
\begin {eqnarray}
k_{RE,Babich}=\frac{N_{RE}(t_{z_f=3\ \textnormal{\tiny
cm}})-N_{RE}(t_{z_f=2.9\ \textnormal{\tiny cm}})}{0.1\ \textnormal{cm}}. \label {runaway_5}
\end {eqnarray}
This definition includes the number of runaway electron produced between 2.9 cm and 3.0 cm which
is close to the end of their simulations. Since previous simulations
and our simulations behave slightly differently, we compare $k_{RE,Babich}$
with $\max\limits_{t}\left(k_{RE}\right)$ calculated from our simulations. $k_{RE,Babich}$
reaches its maximum for $n_{pre,0}=10^{12}$ m$^{-3}$ and amounts to $2\cdot
10^9$ m$^{-1}$ whereas $\max\limits_{t}\left(k_{RE}\right)$ in our simulations reaches
its maximum for $n_{pre,0}=10^{11}$ m$^{-3}$ and amounts to $\approx 2.12\cdot 10^9$ m$^{-1}$.
Thus, we find a good agreement between these two maxima within one order of
magnitude of $n_{pre,0}$. Additionally, we here confirm the findings by
Babich et al. \cite{babich_2015} that the number of runaway electrons is
enhanced if streamers move in preionized air. We also observe both with Monte Carlo or fluid simulations that the runaway
rate per unit length decreases for smaller and larger preionization densities $n_{pre,0}$.

Figure \ref{runaway_2.fig} b) shows that similarly to the maximum runaway rate, the
maximum electron energy increases until $n_{pre,0}\approx 10^{11}$ m$^{-3}$
and decreases for higher levels of preionization. Whereas the maximum
electron energy in non-ionized air is approx. 250 eV, the maximum electron
energy for $n_{pre,0}=10^{11}$ m$^{-3}$ amounts to a tely of 17 keV and decreases to approximately 1 keV for
$n_{pre,0}=10^{13}$ m$^{-3}$ which is sufficiently high for electrons to overcome friction and initiate a relativistic runaway
electron avalanche \cite{gurevich_1992}.

\subsection {The effect of air perturbation on streamer velocities and on
runaway electrons} \label{perturb_effect.sec}
After we have discussed the sole effect of preionization, we now focus on the effect of air perturbations coinciding with the preionization of
air as it might occur in the complex streamer corona in the proximity of
lightning leaders. 

Figure \ref{veloc_1.fig} c) and d) show the mean velocity of the negative front
between $z=4$ cm and $z=6.5$ cm and of the positive front between $z=4$ cm
and $z=1.5$ cm as a function of $n_{pre,0}$. It is seen that for a fixed level of
air perturbation the velocity increases with the
preionization level, which is consistent with what we have observed in
panels a) and b). However, the dominant
contribution in changing the velocity is the air perturbation. Whereas
the mean velocity in non-ionized air is $\approx 5$ mm ns$^{-1}$ for
$\xi=0$, it is $\approx 17$ mm ns$^{-1}$ for $\xi=0.5$, thus a factor of $\approx
3$ larger. This is a consequence of the air density being reduced in the
proximity of the symmetry axis where the density (\ref{dens.1}) of the initial electron-ion
patch is largest.
This is equivalent to the reduced electric field $E/n_{air}(r)$ being enhanced at the location of highest electron density facilitating
the acceleration of streamer fronts parallel to the ambient electric field. For large values of preionization,
the distinction of the streamer velocities becomes even more prominent. For
$n_{pre,0}=10^{13}$ m$^{-3}$ and $\xi=0$, the mean streamer velocity is
$\approx 7$ mm ns$^{-1}$ whereas it is $\approx 27$ mm ns$^{-1}$ for the
same $n_{pre,0}$ and $\xi=0.5$, thus
a factor of $\approx 4$ larger. Thus, the effects of
both preionization and air perturbations increase the streamer
velocity, hence streamers move more significantly when both effects are
combined.

Figure \ref{runaway_2.fig} a) shows $\max\limits_{t}\left(k_{RE}\right)$, as a function of
$n_{pre,0}$ for different levels of air perturbation. 
It shows that in non-ionized air the number of
runaway electrons increases with the level of air perturbation. As for the
streamer velocities, the increase of the number of runaway electrons
results from the enhanced reduced electric field in the vicinity of the
symmetry axis (see \cite{koehn_2018b}). If the effects of air perturbation
and of preionization are combined, we observe a reversed trend compared to
in uniform air. Instead of increasing, the number of runaway electrons in
perturbed air decreases with the preionization level. For $n_{pre,0}\approx
10^{10}$ m$^{-3}$, $\max\limits_{t}\left(k_{RE}\right)$ is approximately $10^{9}$
m$^{-1}$ for all levels air perturbation; for larger $n_{pre,0}$, the
number of runaway electrons in non-perturbed air exceeds that in
perturbed air. Although the runaway rate decreases for $n_{pre,0}=10^{13}$ m$^{-3}$,
$\max\limits_{t}\left(k_{RE}\right)$ is some orders of magnitude
higher in perturbed air than in uniform and non-ionized air. Ultimately, the number of runaway
electrons is highest for $\xi=0.5$ in non-ionized air and for $n_{pre,0}\approx
10^{11}$ m$^{-3}$ in uniform air with $\xi=0$.

Figure \ref{runaway_2.fig} b) compares the maximum electron energy in
perturbed air with that in non-perturbed air for
$n_{pre,0}$. For a perturbation level of
$\xi=0.25$, the maximum electron energy varies from approximately 4 keV to
13 keV for $n_{pre,0}<10^{11}$ m$^{-3}$ which is
one order of magnitude larger than in uniform air and high enough to start
runaway electron avalanches. Above $n_{pre,0}\gtrsim 10^{11}$ m$^{-3}$, the maximum electron energy
varies between approximately 13 keV and 1 keV which is 
comparable to $\max\limits_{t}\left(E_{kin}\right)$ in non-perturbed air.

For $\xi=0.5$, the maximum electron energy is highest and varies between
200 keV and 25 keV which is significant enough to trigger a relativistic avalanche
for any level of preionization. In contrast to $\xi=0$ and $\xi=0.25$, there
is not a distinct maximum between $n_{pre,0}=10^{11}$ and $10^{12}$
m$^{-3}$.

Conclusively, we identify three regimes: i) For air perturbations below
$\xi=0.5$ and for preionization levels below $n_{pre,0}=10^{11}$ m$^{-3}$,
the air perturbation determines the maximum electron energy whereas ii) for
$n_{pre,0}\gtrsim 10^{11}$ m$^{-3}$ the influence of the air perturbation is negligible and
the maximum electron energy is determined by $n_{pre,0}$; iii) For
air perturbations as large as $\xi=0.5$, the maximum electron energy is
mainly determined by the air perturbation with minor effect of the
 preionization level.

\subsection {The streamer velocity and the production of runaway electrons
in subbreakdown fields} \label{sub.sec}
\begin {figure}
\begin {center}
\begin {tabular}{cc}
\includegraphics [scale=0.56] {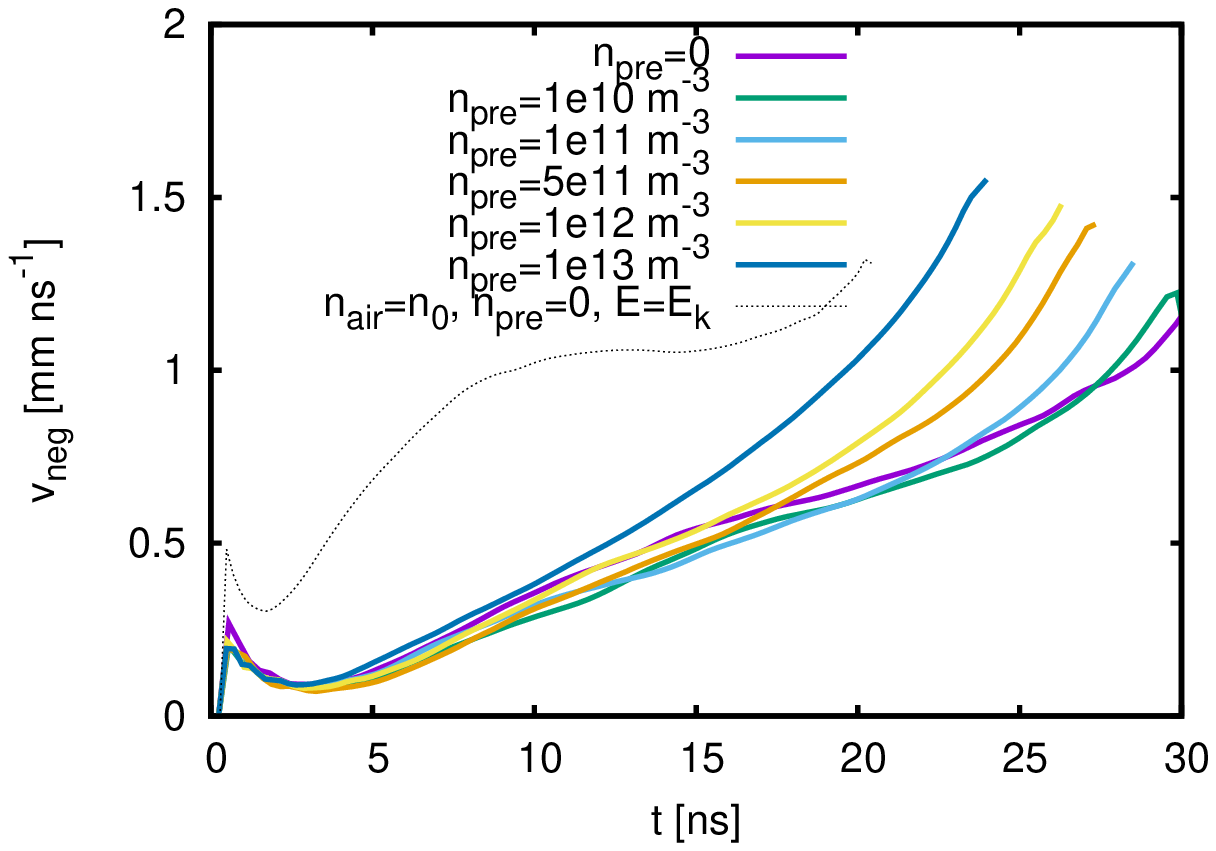} &
\includegraphics [scale=0.56] {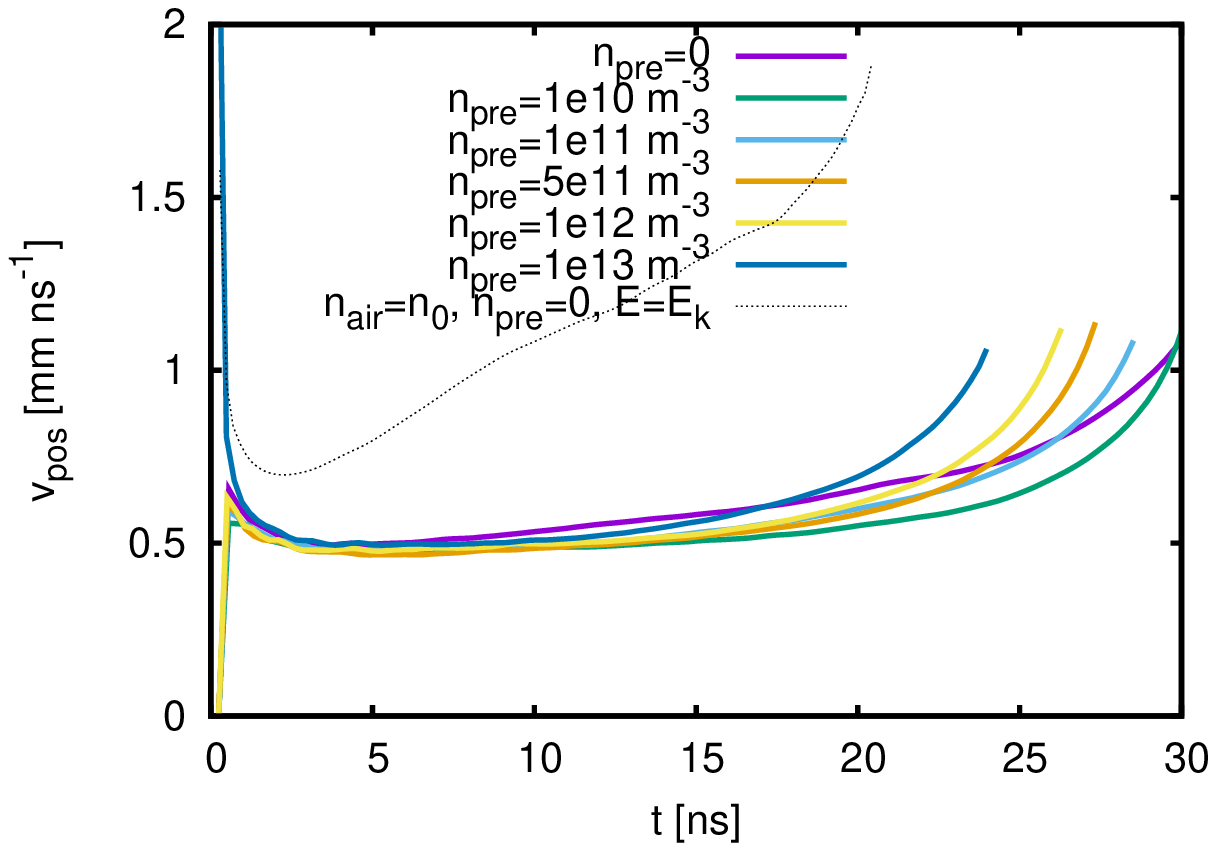} \\
a) $v_{neg}(t)$ & b) $v_{pos}(t)$
\end {tabular}
\end {center}
\caption {The streamer velocities of the negative (a) and positive (b) front as
a function of time for $\xi=0.5$, $E_{amb}=0.5E_k$ and different levels
$n_{pre,0}$ of preionization. For comparison, the dotted line shows the streamer
velocities in non-ionized and uniform air in an ambient field of $E_k$.} \label {veloc_2.fig}
\end {figure}

\begin {figure}
\begin {center}
\begin {tabular}{ccc}
\includegraphics [scale=0.3] {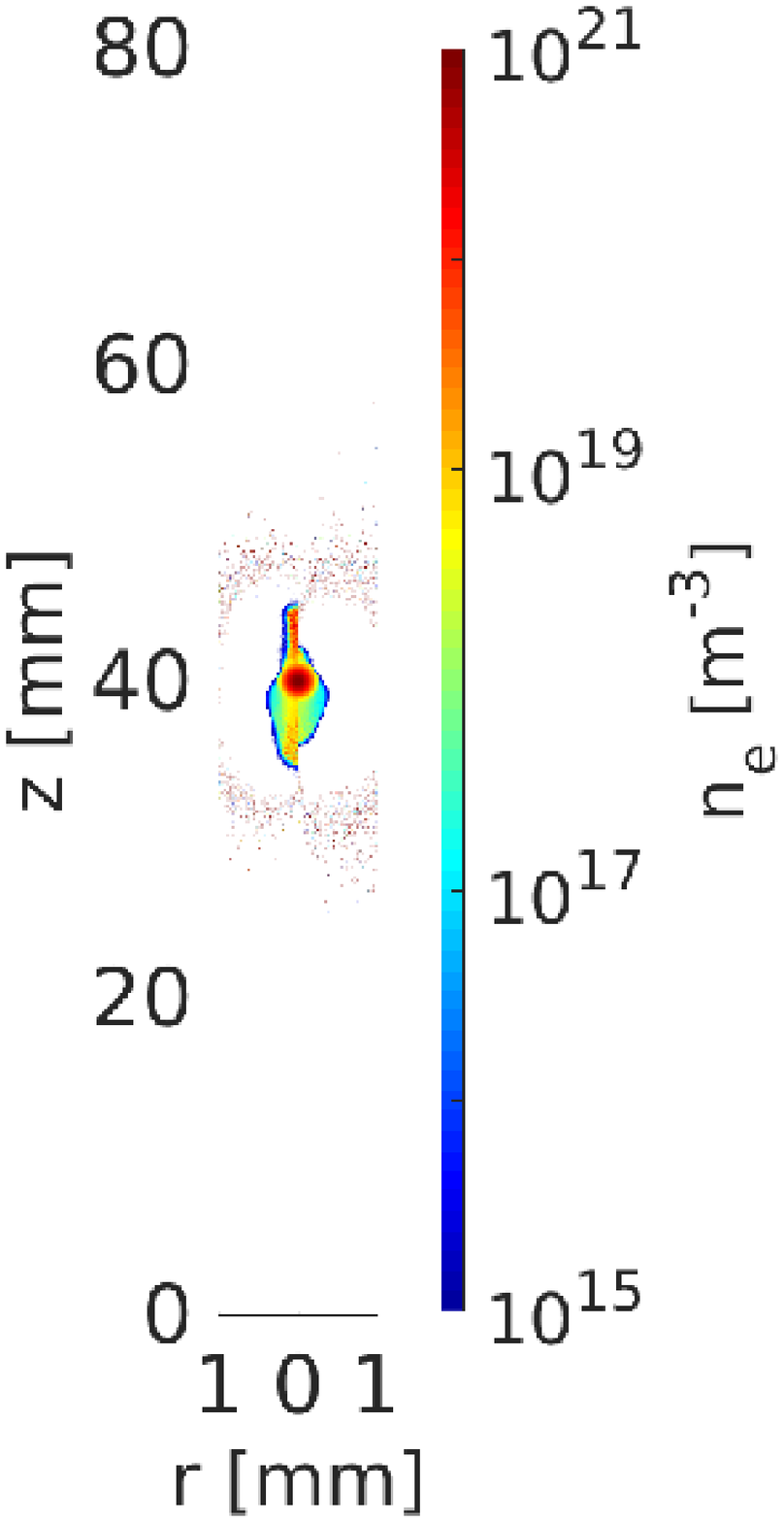} &
\includegraphics [scale=0.3] {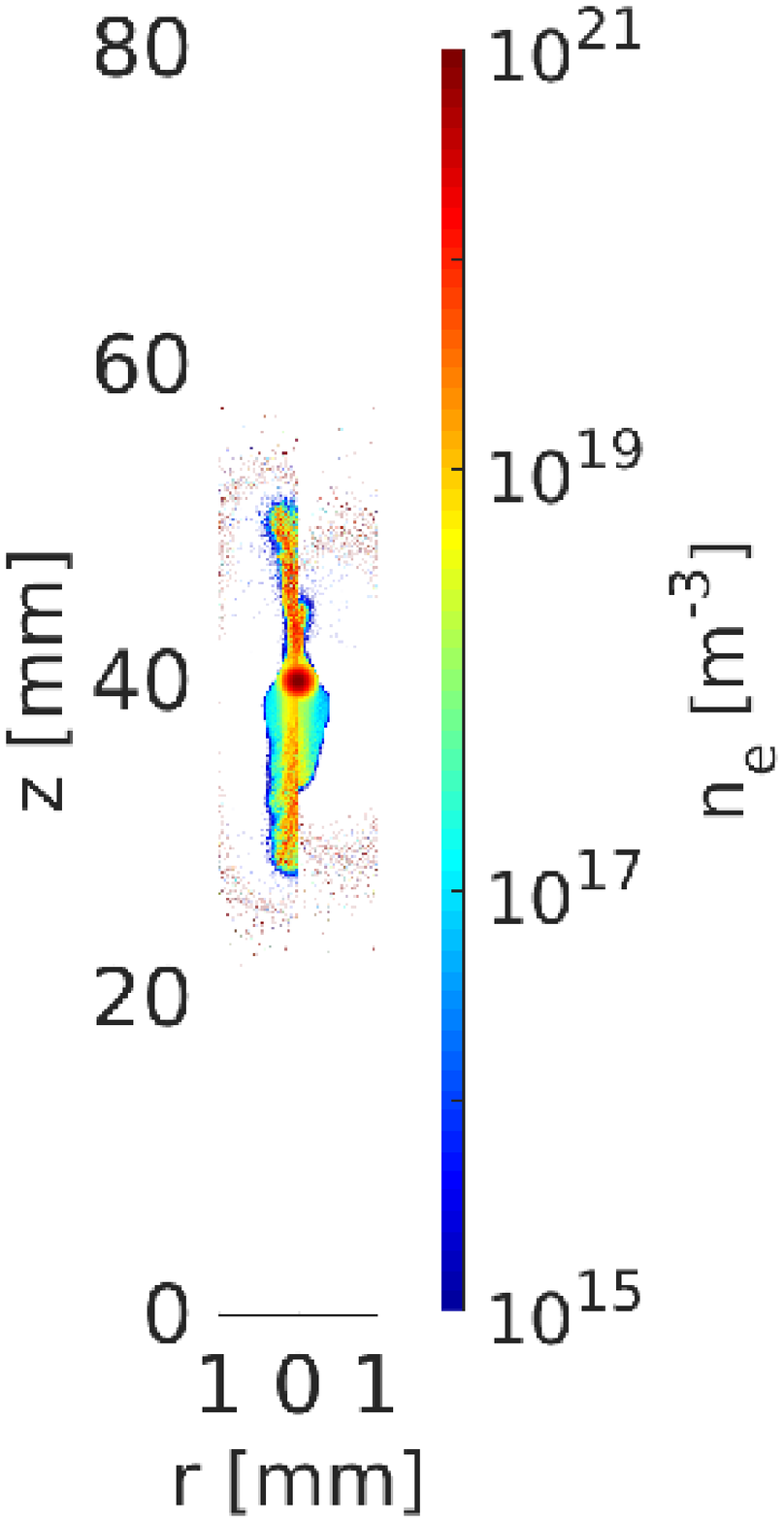} &
\includegraphics [scale=0.3] {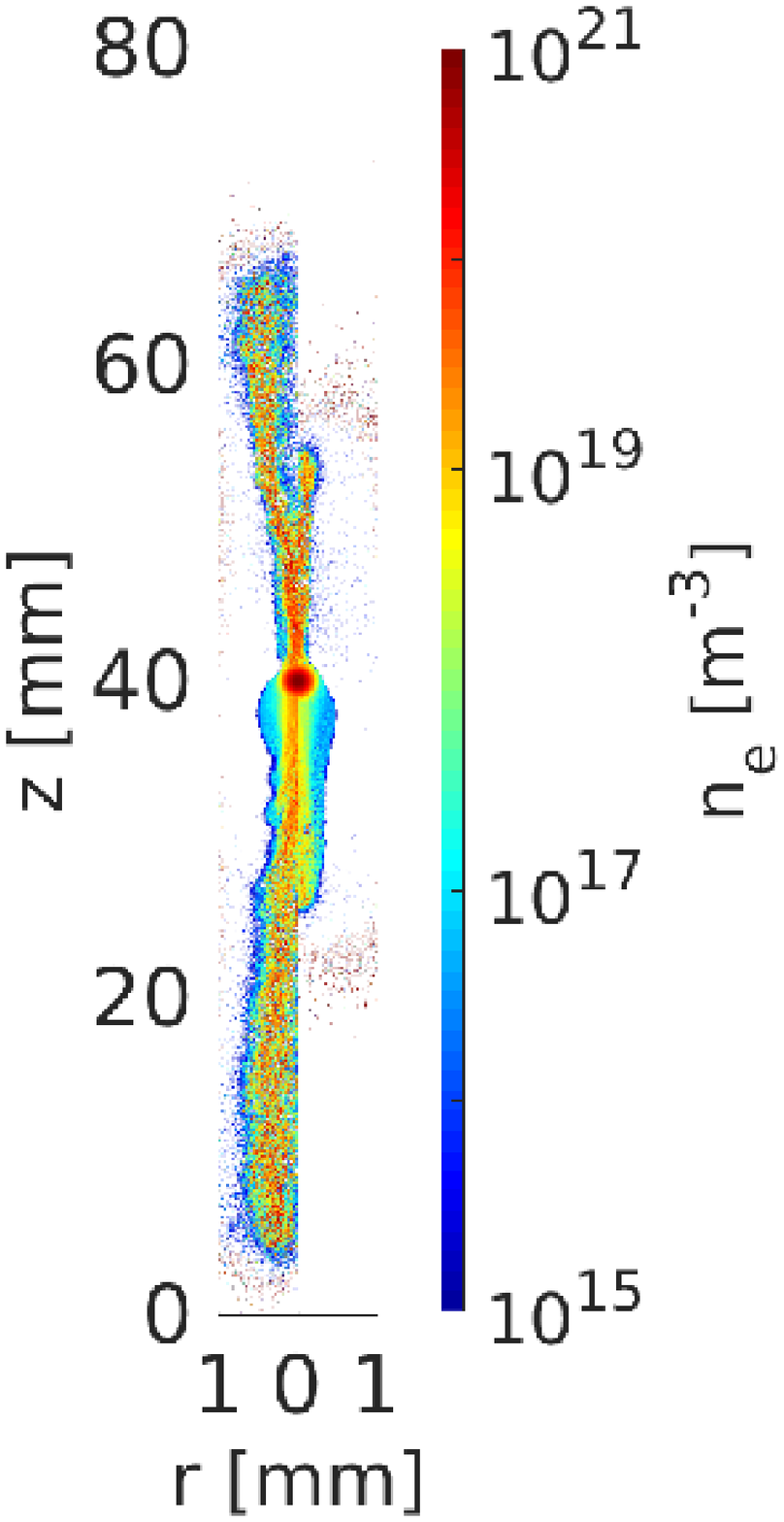} \\
$t=5$ ns & $t=10$ ns & $t=20$ ns
\end {tabular}
\caption {The electron density (first row) in non-ionized air with $E_{amb}=E_k$ and $\xi=0$
(left half of each panel) and with $E_{amb}=0.5E_k$ and $\xi=0.5$ (right
half).} \label {dens_2.fig}
\end {center}
\end {figure}

So far, we have discussed the streamer velocity and the effect of
preionizing and perturbing air on the production of
runaway electrons in an ambient field of
$1.56E_k$. However, the multitude of streamers adjacent to leader stepping
also effects the electric field distribution in the corona.
As an example, we therefore now turn to the production of runaway electrons in a
subbreakdown field of $0.5E_k$. Note that the value of this field refers
to the electric field strengths in uniform air. Thus, $E_{amb}=0.5E_k$ means 0.5
times the breakdown field in uniform air with density $n_0$. Hence, if
$\xi=0.5$, a field of $0.5E_k$ is equivalent to the breakdown
field for $r=0$ and is decreasing for $r>0$. For $\xi<0.5$ the electric field
strength would be below the breakdown field value in the whole simulation domain and
therefore we only consider $\xi=0.5$ and $E_{amb}=0.5E_k$.

For this particular set-up, Figure \ref{veloc_2.fig} shows the streamer velocities at the negative (a)
and positive (b) front as a function of time for different levels of
preionization. We observe a similar dependency
on preionization as in $E_{amb}=1.56E_k$.
For time steps $\lesssim 15-20$ ns, the
preionization effect is negligible; for larger time steps,
streamers accelerate more efficiently when wave fronts grow
into preionized air with larger $n_{pre,0}$. Such waves create more space
charges, thus induce higher self-consistent electric fields and lead to more
efficient streamer acceleration.

For comparison, the dotted line shows the front
velocity in non-perturbed air with $n_{pre,0}\equiv 0$, $\xi\equiv 0$ and
$E_{amb}=E_k$. This comparison reveals that the streamer fronts move significantly slower
for $\xi=0.5$ and $E_{amb}=0.5E_k$ than for $\xi=0$ and $E_{amb}=E_k$
because of the non-uniformity of the air perturbation. Only at
the symmetry axis where $n_{air}=0.5n_0$ is the reduced electric field
$E/n_{air}$ comparably large as $E_k/n_0$ in non-perturbed air. Since the reduced electric field
decreases with $r$, the streamer motion is damped for $r>0$,
and the streamers move slower in unperturbed air.

This effect of air perturbations on the reduced electric field is
visualized in Figure \ref{dens_2.fig}. It shows the electron density
for $\xi=0$ and $E_{amb}=E_k$ and for $\xi=0.5$
and $E_{amb}=0.5E_k$. It shows that after 5 ns
the streamer in uniform air moves faster and is thicker and more diffuse than in perturbed
air. The reason for this is again that the reduced electric field
in perturbed air decreases as a function of $r$ and thus generates a quenching effect on the
streamer in $E_{amb}=0.5E_k$. 


Panels a) and b) of Fig. \ref{runaway_2.fig} compare the maximum runaway production rate and the maximum electron
energy for $\xi=0.5$ and $E_{amb}=0.5E_k$ with the rates and energies of
electrons from streamers in $E_{amb}=1.56E_k$. For $n_{pre,0}\lesssim
10^{12}$ m$^{-3}$ the runaway rates and the maximum electron energy are
smaller than in uniform air. For $n_{pre,0}\gtrsim 10^{12}$
m$^{-3}$, however, $\max\limits_{t}\left(k_{RE}\right)$ becomes comparable to the
one in uniform air in $1.56E_k$. The maximum electron
energy in this set-up increases with $n_{pre,0}$ and reaches
approximately 3 keV which allows the formation of runaway electron avalanches in subbreakdown fields.

\subsection {Variation of the channel radius} \label {radius.sec}
\begin {figure}
\begin {center}
\begin {tabular}{cc}
\includegraphics [scale=0.56] {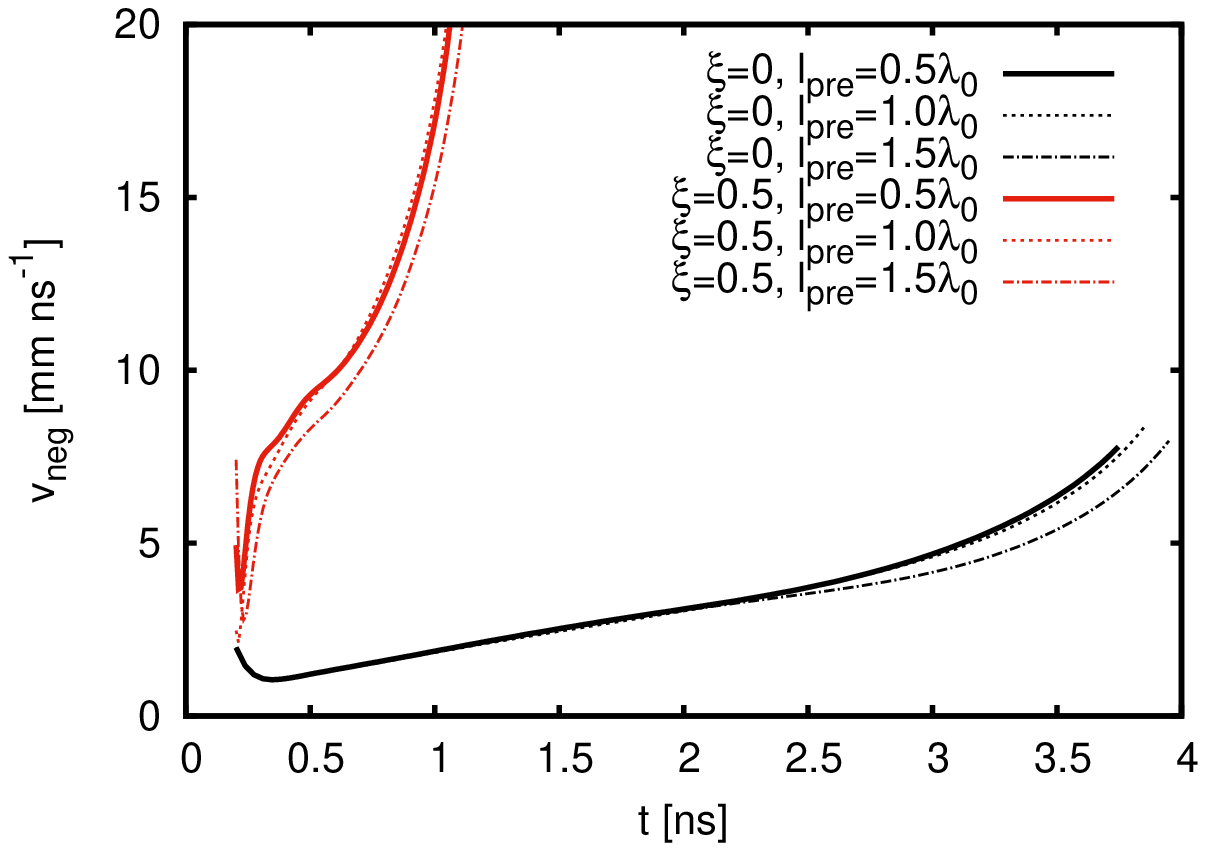} &
\includegraphics [scale=0.56] {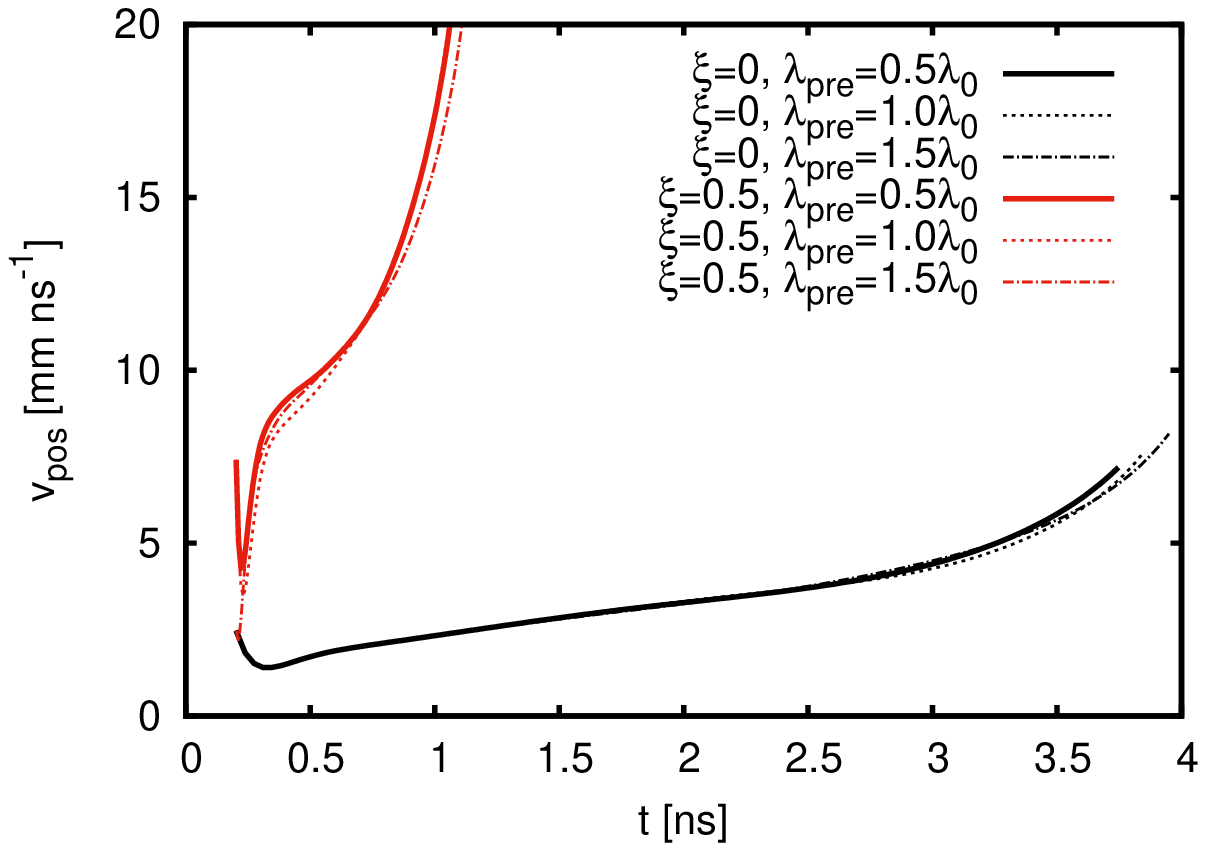} \\
a) $v_{neg}(t)$ & b) $v_{pos}(t)$
\end {tabular}
\end {center}
\caption {The streamer velocities for $n_{pre,0}=10^{11}$ m$^{-3}$ at the negative (a) and positive (b) front
for different $\xi$ and different $\lambda_{pre}$.} \label {veloc_3.fig}
\end {figure}

Whereas the previous results were obtained for $\lambda_{pre}=\lambda_0$, we
now discuss how the streamer velocities and the number of runaway electrons change for
$\lambda_{pre}=0.5\lambda_0$ and for $\lambda_{pre}=1.5\lambda_0$.

Figure \ref{veloc_3.fig} compares the front velocities for different channel
radii $\lambda_{pre}$. Note that the absence of any preionization is equivalent to
$\lambda_{pre}\rightarrow 0$ and the presence of uniform background ionization in
the complete simulation domain is equivalent to
$\lambda_{pre}\rightarrow\infty$. Figure \ref{veloc_3.fig} shows that in non-perturbed air, both the positive and
the negative streamer front move faster for small channel radii and slower
for large channel radii. In contrast, for $\xi=0.5$, there is no significant
difference between $\lambda_{pre}=0.5\lambda_0$ and $\lambda_{pre}=\lambda_0$;
yet, for a wide channel with $\lambda_{pre}=1.5\lambda_0$ the streamer moves slower than in small
channels. This is consistent with modelling results of streamer heads under
different levels of uniform background ionization \cite{wormeester_2010}
which have shown that streamers move faster in the absence of any background
ionization than in the presence of uniform background ionization.

Panels c) and d) of Figure \ref{runaway_2.fig} compare
$\max\limits_{t}\left(k_{RE}\right)$ and
$\max\limits_{t}\left(E_{kin}\right)$ for different $n_{pre,0}$ and
$\lambda_{pre}$. Note that there is no channel for $n_{pre,0}=0$,
thus we only use the values for $n_{pre,0}=0$ as comparison. These panels show that for $\lambda_{pre}\not=\lambda_0$, the maximum
generation rate of runaway electrons is slightly higher than in
preionized air with $\lambda_{pre}=\lambda_0$. Still, the overall
trend is the same regardless of $\lambda_{pre}$. In non-perturbed air, the
maximum generation rate $\max\limits_{t}\left(k_{RE}\right)$ of runaway electrons
reaches its maximum for $\lambda_{pre}\not=\lambda_{0}$ at $n_{pre,0}=10^{10}$ m$^{-3}$, and decreases for
larger preionization levels. In perturbed air, $\max\left(k_{RE}\right)$
is maximal for $n_{pre,0}=0$ and decreases for $n_{pre,0}>0$. 

\section {Discussion and conclusion} \label {concl.sec}
\begin {table}
\begin {tabular}{c|c|c|c}
\backslashbox{$n_{pre,0}$ [m$^{-3}$]}{$\xi$} & 0 & 0.25 & 0.5\\
\hline
0 & $1.56\cdot 10^{8}$ & $6.33\cdot 10^{13}$ & $3.32\cdot 10^{17}$\\
$10^{10}$ & $1.55\cdot 10^{14}$ & $2.37\cdot 10^{14}$ & $2.98\cdot 10^{14}$\\
$10^{11}$ & $6.61\cdot 10^{14}$ & $2.24\cdot 10^{12}$ & $1.04\cdot 10^{14}$\\
$5\cdot 10^{11}$ & $1.72\cdot 10^{13}$ & $5.96\cdot 10^{11}$ & $5.96\cdot 10^{11}$\\
$10^{12}$ & $7.03\cdot 10^{12}$ & $1.52\cdot 10^{12}$ & $1.04\cdot 10^{11}$\\
$10^{13}$ & $5.3097\cdot 10^{12}$ & $5.46\cdot 10^{10}$ & $3.87\cdot 10^{10}$
\end {tabular}
\caption {The rate of runaway electrons per unit time [s$^{-1}$]
for different levels of preionization and air perturbation} \label{runaway_2.tab}
\end {table}

We have discussed how the amount of preionization $n_{pre,0}$, the width $\lambda_{pre}$ of the
preionized channel and the air perturbation level $\xi$
adjacent to leader stepping influence the streamer velocities, the maximum electron energy and production
rate of runaway electrons in streamer discharges and thus
affect the production rate of X-ray bursts and terrestrial gamma-ray flashes
after the leader stepping.

In all considered cases, above and below the classical breakdown field, we
have seen that increasing both the level of preionization and of air
perturbation increases the streamer velocities at the positive and
negative fronts. When increasing the amount of preionization, initially
there is no difference in the streamer velocities before streamers in highly
ionized channels begin to accelerate more prominently than in less ionized
air.  This is due to the enhanced electric field induced by the elevated
amount of space charges produced by streamers growing into channels with
high preionization.  However, the streamer velocity is primarily affected by
air perturbations since the electron motion and hence the streamer
development are determined by the reduced electric field $E/n_{air}$.
In addition, the width of the preionized channel
has a marginal effect on the streamer velocity: Thinner channels
accelerate more significantly than thicker channels.

Our simulations have shown that in the absence of any air perturbations, the
generation of runaway electrons increases with $n_{pre,0}$ up to
$n_{pre,0}\approx 10^{11}$ m$^{-3}$ and decreases for higher preionization since the additional space charges shields the
electric field reducing the electron acceleration and thus the production of
runaway electrons.

Enabling air perturbations in non-ionized air increases the
runaway electron production rate per unit length which is consistent to
previous simulations \cite{koehn_2018b}. However, increasing $n_{pre,0}$
decreases the runaway electron production rate in contrast to increasing
$n_{pre,0}$ in uniform air; yet these rates are larger than in non-ionized and uniform air.

Preionization and air perturbations also allow for the production of runaway
electrons below the classical breakdown field. In a field of
$E_{amb}=0.5E_k$, 50\% air perturbation and preionization levels larger than
$10^{12}$ m$^{-3}$, the production rate of runaway electrons lies within one
order of magnitude of the production rate of runaway electrons
in $1.56E_k$.

Together with these runaway production rates, the maximum electron energies
vary from some keV in non-perturbed  and ionized air up to hundreds of keV in
perturbed air. Under these circumstances, the electron energies are
sufficiently high to create secondary relativistic runaway electron avalanches.

For the cases considered, Table \ref{runaway_2.tab} summarizes the production rate per unit time
defined as $N_{RE}(t_{max})/t_{max}$ where $N_{RE}(t_{max})$ is the total
number of runaway electrons at the end of the simulation $t_{max}$. In
non-perturbed and preionized air, this rate varies between $\approx 10^{14}$
and $10^{12}$ runaway electrons per second. In perturbed air with $\xi=0.25$
and $\xi=0.5$ and $n_{pre,0}<10^{11}$ m$^{-3}$ these
rates vary between $10^{12}$ s$^{-1}$ and $10^{17}$ s$^{-1}$.
Measurements by Schaal et al. \cite{schaal_2012} have revealed that the rate of
energetic electrons producing X-rays adjacent to lightning discharges varies
between $10^{12}$ and $10^{17}$ s$^{-1}$ which can be explained by the
scenarios discussed in the present study.

Celestin and Pasko \cite{celestin_2011} estimate that the streamer corona
in the vicinity of a leader tip consists of approximately $10^6$ streamers.
Hence, applying the runaway rates calculated for one streamer, we estimate the maximum rate of energetic electrons
in preionized or perturbed air to lie approximately between $10^{16}$
s$^{-1}$ and $10^{23}$ s$^{-1}$ for the whole streamer zone. Note, however,
that such a multitude of streamers
influences the properties of each individual streamer through streamer collisions
\cite{ihaddadene_2015,koehn_2017b,luque_2017} as well as through
ionizing \cite{nijdam_2011} or perturbing ambient air \cite{marode_1979,kacem_2013}. Hence, the
runaway rate of $10^{16}-10^{23}$ s$^{-1}$ can only be an upper limit of the real
value of energetic electrons emitted by the whole streamer zone.

Yet, our findings are in agreement with observations.
Thus, the amount of preionization and air perturbation established by
preceding streamers adjacent to lightning leaders is
sufficient to create energetic electrons, significantly multitudinous to
contribute to the emission of X-rays and gamma-rays
in the proximity of lightning leaders in numbers sufficient to account for
terrestrial gamma-ray flashes.

\noindent
{\bf Acknowledgments:} This project has received funding from the European Unions
Horizon 2020 research and innovation programme under the Marie Sklodowska-Curie grant
agreement 722337. The simulations have been performed on the Bridges at PSC
and the Comet at SDSC which are supported by the NSF.

\end {document}